\documentclass[11pt]{article}
\usepackage{style}

\title{(Mis)information diffusion and the financial market
\thanks{This work has received funding from the European Union’s Horizon 2020 research and
innovation programme under the Marie Skłodowska-Curie grant agreement No
956107, “Economic Policy in Complex Environments (EPOC)”.}
}

\author{Tommaso Di Francesco\\
    \href{mailto:t.difrancesco@uva.nl}{\texttt{t.difrancesco@uva.nl}}
    \\ \texttt{University of Amsterdam}
    \\ \texttt{Ca'Foscari University of Venice}  
\and Daniel Torren Peraire\\
    \texttt{Universitat Autònoma de Barcelona}
    \\ \texttt{Ca'Foscari University of Venice} 
    }
    
\date{\today}

\begin{document}
{\setstretch{.8}
\maketitle
\centering

\begin{abstract}

This paper investigates the interplay between information diffusion in social networks and its impact on financial markets
with an Agent-Based Model (ABM).  Agents receive and exchange information about an observable stochastic component of the dividend process
of a risky asset à la \cite{Grossman1980}.
A small proportion of the network has access to a private signal about the component, which can be clean (information)
or distorted (misinformation). Other agents are uninformed and can receive information only from their peers.
All agents are Bayesian, adjusting their beliefs according to the confidence they have in the source of information.
We examine, by means of simulations, how information diffuses in the network and provide a framework to account for delayed absorption 
of shocks, that are not immediately priced as predicted by classical financial models. We investigate the effect of the network topology on the
resulting asset price and evaluate under which condition misinformation diffusion can make the market more inefficient.

\noindent

\end{abstract}
}

\section{Introduction}

Financial markets exhibit empirical regularities that are challenging to capture by relying on the canonical assumption of a represe agent endowed with Full Information Rational Expectation (FIRE) \citep{Muth1961}.
Asset returns display skewed distribution with fat tails and while the Efficient Market Hypothesis \citep{Fama1970} suggests that information is immediately reflected in prices, 
empirical evidence seems to point in the direction of some frictions in its incorporation \citep{Huberman2001,Vozlyublennaia2014}.

A natural starting point in addressing these findings is that of releasing the assumption of a representative agent, and a large body of literature has done so by incorporating agents’ heterogeneity in otherwise standard asset pricing models. 
There are mainly two ways in which this heterogeneity can be modelled. 
The first is assuming that not all agents have rational expectations. 
Theoretical works in this setting have shown how \textit{boundedly rational} agents can survive in the market and how their interaction with rational agents can lead to complex dynamics \citep{Chiarella1992, Brock1997, Brock1998, Lux1998}.

The second relaxation focused on the full information part of the assumption and has been the focus of early research
on information acquisition and processing in financial markets \citep{Kyle1985, Grossman1980, Barberis1998}.
While this literature has focused on the effect of information frictions, it has not fully explored the way in which information is diffused in the market.

In this paper, we construct an Agent-Based Model (ABM) of financial agents connected through a social network to study the effect of information and misinformation diffusion on asset prices. 
ABMs have been widely used in finance and economics, since they allow modeling complex interactions among agents and capturing emergent phenomena that are difficult to predict from individual behavior alone \citep{Dieci2018, Axtell2023}.
We introduce parsimonious relaxations to the FIRE assumption in two dimensions: heterogeneous access to information and delayed transmission of news. 
Only a fraction of agents has access to direct information about the fundamental value of the asset, while the rest of the network can only receive information from their peers.
We then introduce a new social learning mechanism that allows agents to update their beliefs in a Bayesian way while incorporating a behavioral component. 
These modifications, based on empirical evidence, allow us to replicate stylized facts of financial markets without resorting to stronger forms of bounded rationality such as zero-intelligence or purely backward-looking agents. 
Conditional on their heterogeneous information sets, agents are perfectly forward-looking.
To discipline the model, we calibrate it to match moments of the empirical return distribution of TESLA stocks. 
We use the calibrated model to offer an explanation of the delayed absorption of shocks in the market and to investigate the effect of (mis)information diffusion, in different network topologies, on market efficiency.

\subsection{Related literature and Contribution}

Our paper merges insights from two main literatures: one focusing on information diffusion in social networks and one on the effect of information frictions on financial markets. 
In the former, since the seminal work of \cite{Degroot1974} on consensus reaching, scholars have proposed multiple mechanisms of belief updating. 
\cite{Gale2003} are among the first to introduce a network component in the social learning literature.
The presence of an explicit structure leads to the question of whether agents should consider it when making decisions. 
\cite{DeMarzo2003} posit the notion that the information an agent obtains from the network may exhibit bias, since other agents within the network might derive their beliefs from the agent's own beliefs.
After presenting our model, we argue that in our work this problem is not present. 
Moreover, we assume that agents do not think strategically about the network structure.
\cite{Acemoglu2010} introduce a model addressing (mis)information diffusion in network structures where updates occur bilaterally, 
while \cite{Acemoglu2011} and \cite{Kanoria2013} both focus on Bayesian learning. 
\cite{Buechel2015} emphasizes the role of conformity and centrality in shaping the collective wisdom of the network. 
Conversely, \cite{Rusinowska2019} propose a model in which individuals strategically aim to exert influence on others. 
In our context, strategic considerations are absent, and communication is truthful, which differentiates our work from previous literature that considered the effect of access to private information to manipulate markets \cite{Benabou1992}.

We propose a mechanism of belief updating when receiving multiple sources of information simultaneously. 
The mechanism is quite general and nests other social learning models, like the one of \cite{Degroot1974}. 
It relies on Bayesian updating with a behavioral component, since agents construct a time-varying measure of precision for each source of information and use it to derive the posterior mean and variance of the signal of interest. 
The second object is particularly important in our case, as agents are risk-averse and will use this variance to compute their optimal demands.

Information frictions have been documented empirically by multiple sources. 
\cite{Huberman2001} show that the stock prices of a company, CASI Pharmaceuticals, did not incorporate new information for five months. 
They point out that the news was initially released as a research article in the journal Nature, but investors reacted only when a Wall Street Journal article reposted the findings of the original study. 
Behavioral factors can also contribute to information delay. \cite{Vigna2009} provide evidence of limited attention, demonstrating reduced investor responsiveness on Fridays and identifying profitable strategies that exploit such underreactions. 
This delay has been studied theoretically, with a focus on insider trading \citep{Kyle1985, Benabou1992, CollinDufresne2016} or herding behavior in information acquisition \citep{Banerjee1992, Orlean1995, Cont_Bouchaud_2000}.

Finally, a number of previous studies have examined the role of network dynamics in financial markets. 
Most of this work has used networks to model imitation among agents, such as in \cite{Iori2002}. 
\cite{Panchenko2013} build on the \cite{Brock1998} framework, allowing agents to choose among various trading strategies observed within their network. 
\cite{Khashanah2016} develop an ABM in which agents share their optimal holdings with their neighbors. 
\cite{Wu2018} also explore different network topologies within an ABM of financial markets, focusing on how traders switch between fundamentalist and chartist strategies. 
In contrast, \cite{Biondo2020} examines various network structures where agents imitate the discrete trading decisions of their peers, while \cite{Bertella2021} considers scenarios in which agents compare their wealth to that of their neighbors, leading to asset reallocation when their performance lags behind.
The novelty of our model lies in the fact that agents do not imitate each other. As already remarked, all agents are forward-looking, and the network is used only to model information flow.

The rest of the paper is structured as follows: Section \ref{sec:model} introduces the model, focusing on the two blocks
that consittutes the ABM, the financial market and the information diffusion process.
Section \ref{sec:calibration} describes the process we use to obtain a realistic calibration of the model.
Section \ref{sec:numerical_simulations} presents properties of the model by means of numerical simulations.
Section \ref{sec:conclusion} concludes.

\section{Model}\label{sec:model}
\subsection{The financial market}
Consider an economy with I consumers,
indexed by $i = (1,2,\dots, $I$)$. Consumers are infinitely lived, and at the beginning of every period, they receive the same endowment $W_0$.
They have the same utility over the end of period wealth, given by
\begin{equation}\label{Utility}
U(W_t) = -e^{-aW_{t}},
\end{equation}
where $a>0$ is the coefficient of risk aversion.
In order to transfer wealth from the beginning of the period to the end, there are two types of security:
a risk-free and a risky asset. We define $p_t$ as the price of the risky asset at a generic time $t$ and normalize the price of the risk-free asset to 1.
In every period consumers decide how to allocate their initial endowment, choosing between the two possible securities.
At the end of the period they receive profits based on their portfolio, and immediately consume their wealth.
Given their participation in the financial market, throughout the paper we use interchangeably the terms consumers,
investor and agent. Defining $X_{i,t}$ as the consumer's demand for the risky asset and $M_{i,t}$ as the demand for the safe asset,
the allocation choice is subject to the budget constraint
\begin{equation}\label{Budget_constraint}
p_t X_{i,t} + M_{i,t} = W_{0}.
\end{equation}
The risk-free rate is $R > 1$ and the risky asset pays a stochastic payoff which is equal to a dividend claim plus the future price of the asset
\begin{equation}\label{payoff}
y_{t+1} = p_{t+1} + d_{t+1}.
 \end{equation}
The presence of the future price on the right-hand side of equation (\ref{payoff}) ensures a positive feedback mechanism of expectations.
This is a well-established feature of financial markets, as documented among others by \cite{Heemeijer2009}.
There are two stochastic components determining the realization of future dividends
 \begin{equation}
    d_{t+1} = d + \theta_{t+1} + \varepsilon_{t+1},
 \end{equation}
with $\varepsilon_{t+1} \sim \mathcal{N}(0,\sigma^2_\varepsilon)$ being pure unobservable noise.
The stochastic component $\theta_{t+1}$ follows a stationary\footnote{This is a simplifying assumption since it would imply stationary prices. 
A more accurate representation could be given by considering a growth model for dividends like in \cite{Diks2008} and having agents forecast the growth rate.
However, while more realistic, this assumption would not change the main insight of the paper, nor affect the calibration which is based on the model returns.}
AR(1) process, $\theta_{t+1} = \beta \theta_{t} + \eta_{t+1}$ with $\eta_{t+1} \sim \mathcal{N}(0,\sigma^2_\eta)$ and $\beta \in (0,1)$ and is observable by some agents before its actual realisation.
One can think of it as information, in a similar spirit to \cite{Grossman1980} and \cite{Gerotto2019}.
This implies that the underlying fundamental value of the risky asset is stochastic but that some information about it is revealed in advance.
However, this information is not immediately incorporated in the asset price. 
The end of period $t$ wealth for the $i^{th}$ consumer is given by
 \begin{equation}\label{wealth}
     W_{i,t} = R M_{i,t} + y_{t+1} X_{i,t} = R(W_0 - p_t X_{i,t}) + y_{t+1} X_{i,t}.
 \end{equation}
Agents optimize their end of period wealth, which given the normality of ${y_{t+1}}$ is also normal.
The optimization problem is therefore given by

\begin{equation}\label{lognormality}
\max_{\{X_{i,t}\}} \left(-exp\left\{-a\Tilde{\mathbb{E}}_{i,t}(W_{i,t}) + \frac{a^2}{2}\Tilde{\mathbb{V}}_{i,t}(W_{i,t})\right\}\right).
\end{equation}
Using equation (\ref{wealth}) and solving for the optimal choice of risky asset yields
\begin{equation}\label{demand}
X_{i,t} = \frac{\Tilde{\mathbb{E}}_{i,t}(y_{t+1}) - Rp_t}{a\Tilde{\mathbb{V}}_{i,t}(y_{t+1})}.
\end{equation}
The notation $\Tilde{\mathbb{E}}_{i,t}$ and $\Tilde{\mathbb{V}}_{i,t}$ is used to represent subjective expected value and subjective variance for agent $i$.
It is short notation $\mathbb{E}(\cdot | \mathcal{I}_{i,t})$ and $\mathbb{V}(\cdot | \mathcal{I}_{i,t})$, where $\mathcal{I}_{i,t}$ is the information set of agent $i$ at time $t$, that is before the realization of $\theta_{t+1}.$ 
We set net supply of outside share of the risky asset equal to 0 and use the market clearing condition, imposing supply equal to aggregate demand, to obtain an implicit pricing equation
\begin{equation}\label{eq:price}
    \sum_{i=1}^I {X_{i,t}} = \sum_{i=1}^I \left( \frac{\Tilde{\mathbb{E}}_{i,t}(y_{t+1}) - Rp_t}{ \Tilde{\mathbb{V}}_{i,t}(y_{t+1})}  \right) = 0.
\end{equation}
All agents in the model are assumed to be forward-looking in evaluating the asset price.
They expect the asset price to be its fundamental value, which is determined by the present discounted value of the stream of future dividends.
This implies that in the model there is only a minimal deviation from rationality, given by the asymmetry in information.
This is a crucial aspect and one of the main features that distinguish our paper from other prominent works in the literature.
\cite{Chiarella1992}, \cite{Brock1998} and \cite{Lux1998} among the others, focus on the coexistence of (rational) fundamentalists and some type of boundedly rational backward-looking agent.
The most common type of bounded rationality for the latter is associated with technical trading rules and the literature has usually labeled them as chartists or trend followers \citep{Day1990, Tramontana2010, Tramontana2013, Anufriev2020,Gardini2025}.
This interplay has been shown to generate complex dynamics, and in some cases, even chaotic behavior.
A third type of agents which are sometimes included in this framework are sentiment followers \citep{Gardini2022,DiFrancesco2024} and their presence 
can contribute to destibilize the market even further.
By contrast, in our model, all agents act optimally given the information available and if frictions were removed from the information diffusion process, we would end back in the rational expectations setting.
When solving the forward-looking problem, agents in the model assume that other agents behave identically.
This assumption leads them to solve the problem as if they were the representative agent.
In other words, they consider the aggregate behavior of all agents to be equivalent to their own individual behavior. 
With this assumption the pricing equation for each individual is given by
\begin{equation}\label{price_hom}
    p_t  = R^{-1} \Tilde{\mathbb{E}}_{t}\left(y_{t+1}\right),
\end{equation}
in which we omit the subscript $i$ and solving by iterating forward, which is done in section \ref{sec:flp} of the Appendix gives 
\begin{equation}\label{hom_solution}
    p_t = \frac{d}{r} + \frac{\Tilde{\mathbb{E}}_t(\theta_{t+1})}{R - \beta}.
\end{equation}
The first component of equation (\ref{hom_solution}) is the usual discounted value of future expected dividends: without
information, the price would be constant for all time periods.
The second component is specific of our model and imposed by the presence of the observable component of dividend $\theta$.
The next section is devoted to describe the mechanism for which agents receive information about this component.

\subsection{Actual and Perceived Law of Motion}
We now describe the individual beliefs, by distinguishing between the actual law of motion (ALM) of the observable component of dividends $\theta_{t+1}$ 
and the perceived law of motion (PLM) which agents believe to be true. 
We note that the actual process $\theta_{t+1}$ is not publicly available,
and it can not be directly observed even after its realization.
A subset of agents, whom we call informed and misinformed, have access to a private signal which they believe to be the true process.
They have full confidence in their source of information and as soon as they receive this private signal, they update their beliefs to it.
This behavior can be sustained by the fact that they only observe the realization of payoffs.
Errors in their payoff forecasts can be attributed to the idiosyncratic noise $\varepsilon_{t+1}$ and not necessarily to their source of information.
The remaining agents, which we call uninformed, do not have access to the private signal and update their beliefs in a Bayesian way,
which we specify below.
The ALM is given by the AR(1) process
\begin{equation}
    \theta_{t+1} = \beta \theta_t + \eta_{t+1}.
\end{equation}
This implies that the conditional distribution of $\theta_{t+1}$ given the information set $\mathcal{I}_t$ is
\begin{equation}
    \theta_{t+1} | \mathcal{I}_t \sim \mathcal{N}\left(\beta \theta_t, \sigma_{\eta}^2 \right).
\end{equation}
Since we assume that all agents are aware of the structure of the process, we model all prior beliefs as normal distributions.
As anticipated we assign agents to three possible categories. 

\vspace{5mm}

\noindent
\textbf{Informed agents.} \hspace{1mm} These agents receive the true information $\theta_{t+1}$ and base their forecast on it.
One can think that this is due to agents having access to privileged or inside information. 
We prefer to associate this choice with empirical evidence provided by \cite{Huberman2001} and \cite{Peng2006} supporting the idea of different classes of investors with different access to information.
Some agents may possess knowledge to process domain-specific information that other, generalist agents, may lack.
Their PLM coincides with the ALM.
For them we have
\begin{equation}
\theta_{t+1} | \mathcal{I}_t \sim \mathcal{N}\left(\beta \theta_t, \sigma_{\eta}^2 \right).
\end{equation}
\noindent
\textbf{Misinformed agents.} \hspace{1mm} These agents think they have perfect foresight like informed agents, but are actually basing their 
forecast on misinformation. Their PLM is given by

\begin{equation}
    {\theta}_{t+1} = \beta \gamma_t + \sigma_{\nu}: = \gamma_{t+1}, 
\end{equation}

that is they assume the process follows an AR(1),
with same persistence parameter $\beta$ but different noise term $\nu_{t+1} \sim \mathcal{N}(0,\sigma^2_{\nu})$.
This can be thought of as a misinformation shock.
Their conditional distribution is then given by

\begin{equation}
\theta_{t+1} | \mathcal{I}_t \sim \mathcal{N}\left(\beta \gamma_t, \sigma_{\nu}^2 \right). 
\end{equation}

One can think of these traders as noise traders in \cite{DeLong1990} style or akin to sentiment followers.
One could naturally ask why these individuals are trading on noise. As \cite{BLACK1986} puts it,
\textit{``One reason is that they like to do it. Another is that there is so much noise around that they don't know they are trading on noise. They think they are trading on information."}
The latter explanation is what we argue can sustain this behavior in our model.
Although these agents are not evaluating their source of information, they might be unable to detect their bias in the presence of noise.

\vspace{5mm}

\noindent
\textbf{Uninformed agents.} \hspace{1mm} These agents do not have any private signal, but their PLM coincides with the ALM.
The only difference is that they condition their beliefs on their previous period expectation,

\begin{equation}
    \theta_{t+1} |  \mathcal{I}_t  \sim \mathcal{N}\left(\beta \Tilde{\mathbb{E}}_{t-1}(\theta_t) , \sigma^2_{\eta} \right).
\end{equation}
The literature provides ample evidence that not all information is immediately processed by investors upon release. 
The simplest explanation is that of limited attention. 
\cite{Hirshleifer2009} finds that investor reactions to earnings announcement are weaker on days when there are multiple simultaneous 
news releases. On a similar note \cite{Vigna2009} show that investors take more time to process news on Friday.
\cite{Tetlock2011} shows that investors overreact to stale information and \cite{Gilbert2012} demonstrate that this causes mispracing in the market by constructing a profitable trading strategy
exploiting this finding. More recently \cite{Blankespoor2020} show that there seems to exist some ``disclosure processing costs" 
that would make disclosures not public information, as usually defined, but a form of private information.

\subsection{Information diffusion}\label{sec:information_diffusion}
Agents are socially connected and are organized in a network. Each agent represents a node, and nodes are entirely characterized by beliefs
regarding the observable component of dividends $\theta_{t+1}$. 
The network is static. All edges are exogenously determined and time-invariant. The edges represent the flow of information between nodes.
At the beginning of time $t$ agents have normal prior distribution according to their category. 
They then receive new data in two ways.
Informed and misinformed agents observe $\theta_{t+1}$ and $\gamma_{t+1}$ respectively.
Moreover they observe them without noise, so that their posterior beliefs are immediately updated to the realizations of the variables\footnote{This is in a sense an abuse of notation as a Normal distribution with 0 variance is a degenerate distribution with support at the single point $\theta_{t+1}$, known as the Dirac’s delta function. 
However this notation is convenient since it allow us to model the evolution of the beliefs of this category of agents in the general framework.}
\begin{equation}
     \theta_{t+1} |  \theta_{t+1}  \sim \mathcal{N}\left(\theta_{t+1}, 0\right) \quad  \gamma_{t+1} |  \gamma_{t+1}  \sim \mathcal{N}\left(\gamma_{t+1}, 0\right).
\end{equation}
Uninformed agents do not receive any private information, but they can receive information from their peers.
Formally, for agent $i$, we assume that they can observe node $j$ prior mean if an edge exists between node $i$ and node $j$. 
Based on this information they update their beliefs in a Bayesian way but incorporating a behavioral component. 
When an agent is connected to another node in the network, we assume that they construct an
implicit variance evaluating the forecasting error of the node over time. To do so we compute an exponential moving average of the 
forecasting error, given as the squared difference between the last observable payoff and the payoff prediction $(\mu_{j,t-1})$ implied by source $j$,
that is
\begin{equation}\label{sample_var}
    EMA_{j,t} = w \left(y_{t-1} - \left(\frac{dR}{r} + \mu_{j,t-1}\right)\right)^2 + (1-w) EMA_{j,t-1}.
\end{equation}
Then, to map this to a comparable variance of the given source, we multiply the ratio between source $j$ exponential moving average and
their own, with the prior variance

\begin{equation}
\sigma^2_{j,t} = \sigma^2_{\eta} \frac{{EMA_{j,t}}}{{EMA_{i,t}}}.
\end{equation}
In simple terms, if they observe that node $j$ has been more accurate then themselves they attach higher confidence to the information received by that node.
They then update their beliefs by applying the Bayes rule to the case of receiving information
from $K$ different sources and given in the following proposition.

\begin{proposition}[Bayesian Updating of Beliefs]
\label{thm:BA}

Assume agents have a normal prior distribution with parameters $(\mu_0, \sigma^2_0)$ and receive $K$ new information, each with a Normal likelihood $(\mu_k, \sigma^2_k)$,  
$k = 1, 2\dots  K$. Then agents posterior distribution is Normal,
with posterior mean given by:
\begin{equation}\label{post_mean}
\mu_P = \frac{\sum^{K}_{k=0}\left(\mu_k \cdot \left[ A \right]^{\Bar{A}-1}_{k}\right)}{\sum \left[ A \right]^{K}},
\end{equation}
 and posterior variance:
\begin{equation}\label{post_variance}
\sigma^2_P = \frac{\prod^{K}_{j=0}\sigma^2_j}{\sum \left[ A \right]^{\Bar{A}-1}},
\end{equation}
where,
$A = \{\sigma^2_0, \sigma^2_1,\sigma^2_2, \dots \sigma^2_K\}$, $\Bar{A}$ is the cardinality of set $A$. $\left[ A \right]^J$ is the set of all distinct combinations of products of size J from set A,
$\sum \left[ A \right]^J$ sums over all the elements in the set, $\left[ A \right]^J_{k}$ indicates the combination that does not include $\sigma^2_k$.
\end{proposition}
\textit{Proof.} See appendix (\ref{sec:proof_BA}).

In every period $t$ agent $i$ uses this mechanism to update their beliefs and obtain their posterior distribution $\theta_{t+1} \sim \mathcal{N}(\mu_{P,t}, \sigma^2_{P,t})$.
To build intuition consider the simplest case in which an agent is connected only to one other node.
Then the posterior parameters are given by
\begin{equation}
\mu_P = \frac{\mu_0 \sigma^2_1 + \mu_1 \sigma^2_0}{\sigma^2_0 + \sigma^2_1}, \quad \sigma^2_P = \frac{\sigma^2_0 \sigma^2_1}{\sigma^2_0 + \sigma^2_1}.
\end{equation}
Notice that this updating procedure is equivalent to using the Kalman filter to filter out the noise in the signal, as we show in appendix \ref{sec:relationship_kalman_filter}.
If one then moves to $K=2$, hence considering an agent with two connections, posterior parameters are given by

\begin{equation}
    \mu_P = \frac{\mu_0 \sigma^2_1 \sigma^2_2 + \mu_1 \sigma^2_0 \sigma^2_2 + \mu_2 \sigma^2_0 \sigma^2_1}{\sigma^2_1 \sigma^2_2 +  \sigma^2_0 \sigma^2_2 + \sigma^2_0 \sigma^2_1}, \quad \sigma^2_P = \frac{\sigma^2_0 \sigma^2_1 \sigma^2_2 }{\sigma^2_1 \sigma^2_2 +  \sigma^2_0 \sigma^2_2 + \sigma^2_0 \sigma^2_1}.
\end{equation}

The posterior mean can be seen as a weighted average of the prior means the agent has access to, with weight given by ${\left[ A \right]^{\Bar{A}-1}_{j}}/{\sum \left[ A \right]^{\Bar{A}-1}}$.
The noisier the alternatives are, the more an agent will rely on a particular source.
This mechanism can be seen as a particular case of the naive updating proposed in \cite{Golub2010} but the novelty of our approach is that we simultaneously derive the posterior variance. 
This object while of no relevance in most information diffusion works, has a crucial role in our work, given the risk averse behavior of our agents. 
Also different in our case is that the information exchange happens only one time per time step, 
therefore avoiding any potential bias given by repeated information as is the case in \cite{DeMarzo2003}.
Morevoer the only source of heterogeneity is given by the different categories of agents, but within the same class, beliefs are ex-ante homogenous. 
This implies that there are only two innovations or shocks entering the network at each time step. 
The combination of this property with the autoregressive structure of the component agents are interested in, make so that aggregating over correlated information is unavoidable but not deleterious. 
Of particular interest are then the following situations. 
\begin{enumerate}
    \item An agent considers source $k$ to be absolutely certain. Then in our model we have
    \begin{equation}\label{limit_case1_mean}
    \sigma^2_k = 0 \implies  \mu_p = \mu_k,
\end{equation}
\begin{equation}\label{limit_case1_var}
    \sigma^2_k = 0 \implies \sigma^2_p = 0.
\end{equation}

That is, when an agent uses only one source of information and is totally confident in it, 
the posterior mean is equal to the signal, with variance 0.
\item An agent completely disregards source $k$, then 
\begin{equation}\label{limit_case2_mean}
\lim_{\sigma^2_k \to +\infty} \mu_p = \frac{\sum^{K}_{k=0}\left(\mu_k \cdot \left[ B \right]^J_{k}\right)}{\sum \left[ B \right]^{\Bar{B}-1}},
\end{equation}

\begin{equation}\label{limit_case2_var}
    \lim_{\sigma^2_k \to +\infty} \sigma^2_p = \frac{\prod^{K}_{j=0}\sigma^2_j}{\sum \left[ B \right]^{\Bar{B}-1}}.
\end{equation}

where $B = A \setminus \sigma^2_k$.
When agents believe that a source of information is totally unreliable, their posterior mean and variance is equal to the one they would get by omitting the source of information.
\end{enumerate}

\subsection{Payoffs and prices}

With the posterior beliefs regarding the observable component of the dividend \( \theta_{t+1} \), we can compute beliefs regarding future payoffs. First, the subjective expectations are \( \Tilde{\mathbb{E}}_t(\theta_{t+1}) = \theta_{t+1} \) for informed agents, \( \Tilde{\mathbb{E}}_t(\theta_{t+1}) = \gamma_{t+1} \) for misinformed agents, and \( \Tilde{\mathbb{E}}_t(\theta_{t+1}) = \mu_{P,t} \) for uninformed agents. Using the expression for \( p_t \) allows us to compute, (see Appendix \ref{sec:conditional_variance}), the conditional expectation:  

\begin{equation}\label{eq:expected_payoff}
    \Tilde{\mathbb{E}}_t(y_{t+1}) = \frac{dR}{r} + \frac{R \Tilde{\mathbb{E}}_t(\theta_{t+1})}{R-\beta},
\end{equation}  

and the conditional variance:  

\begin{equation}\label{eq:exp_variance}
   \Tilde{\mathbb{V}}_t(y_{t+1}) = \sigma^2_{\varepsilon}
   + \Tilde{\mathbb{V}}_t(\theta_{t+1}) +  \frac{\Tilde{\mathbb{V}}_t(\Tilde{\mathbb{E}}_{t+1}(\theta_{t+2}))}{(R-\beta)^2} + 2\Tilde{\mathbb{COV}}_t \left(\theta_{t+1}, \frac{\Tilde{\mathbb{E}}_{t+1}(\theta_{t+2})}{R-\beta}\right),
\end{equation}  

which clarifies that heterogeneity in beliefs is completely described by \( \theta_{t+1} \). While the expected value of the stochastic payoff is given for every category by equation (\ref{eq:expected_payoff}), the conditional variance is specific for each category. In particular, starting from equation (\ref{eq:exp_variance}), we have:  

\vspace{5mm}

\noindent
\textbf{Informed agents}  

\begin{equation}
\Tilde{\mathbb{V}}_t(y_{t+1}) = \sigma^2_{\varepsilon}
   + \frac{\sigma^2_{\eta}}{(R-\beta)^2},
\end{equation}  

since for these agents, \( \theta_{t+1} \) is not a random variable at time \( t \), and  
\begin{equation}
    \Tilde{\mathbb{V}}_t(\Tilde{\mathbb{E}}_{t+1}(\theta_{t+2})) = \Tilde{\mathbb{V}}_t(\theta_{t+2}) = \sigma^2_{\eta}. 
\end{equation}

\vspace{5mm}

\noindent
\textbf{Misinformed agents}  

Similarly, for misinformed agents, we have:  
\begin{equation}
\Tilde{\mathbb{V}}_t(y_{t+1}) = \sigma^2_{\varepsilon}
   + \frac{\sigma^2_{\nu}}{(R-\beta)^2}.
\end{equation}

\vspace{5mm}

\noindent
\textbf{Uninformed agents}  

Deriving the conditional variance for uninformed agents is more challenging. In particular, the term \( \Tilde{\mathbb{E}}_{t+1}(\theta_{t+2}) \) depends on the network structure and the updating mechanism we have described. This makes the future forecast a random variable, whose distribution depends on the entire network topology, making the derivation analytically intractable. However, since we are deriving the conditional variance of payoffs under the assumption that agents operate as representative agents, we have:  
\begin{equation}
    \Tilde{\mathbb{E}}_{t+1}(\theta_{t+2}) = \beta \Tilde{\mathbb{E}}_{t}(\theta_{t+1}), 
\end{equation}
which is known at time \( t \) and, therefore, not a random variable. Under this assumption, we have:  
\begin{equation}
\Tilde{\mathbb{V}}_t(y_{t+1}) = \sigma^2_{\varepsilon}
   + \Tilde{\mathbb{V}}_t(\theta_{t+1}).
\end{equation}  

\subsection{The role of social learning}

In this section, we highlight the importance of our social learning model by comparing the full model to two simpler cases that are naturally nested within our framework.  

The first case features traders’ heterogeneity but no social learning. In this framework, agents keep their expectations constant at their prior beliefs. Defining \( I \) as the total number of agents, and \( \lambda \) and \( \xi \) as the proportions of informed and misinformed agents respectively, and imposing a prior mean of 0 for uninformed agents, the resulting price evolves according to:  
\begin{equation}
    p_t = \frac{d}{r} + \frac{\lambda \theta_{t+1} + \xi \gamma_{t+1}}{R - \beta}.
\end{equation}  

The second case modifies this baseline by adding perhaps the most notable social learning mechanism, that of \cite{Degroot1974}, which occurs when uninformed agents assign equal weight to all sources of information available. For this case, we consider a fully connected network, allowing us to derive analytical results. In this framework, every uninformed agent receives \( \lambda I \) signals from informed agents, \( \xi I \) signals from misinformed agents, and \( (1 - \lambda - \xi)I \) signals from uninformed agents (including themselves).  

Moreover, every uninformed agent has identical beliefs. In particular, assuming that the homogeneous variance attached to every piece of information is \( \sigma^2 > 0 \), we have for uninformed agents:  
\begin{equation}
   \Tilde{\mathbb{V}}_t(\theta_{t+1}) =  \frac{\sigma^2}{I}.
\end{equation}  

It is worth noting that the posterior variance approaches 0 as the network size approaches infinity. More precisely, the driving factor here is the number of signals each agent receives, which equals the total number of agents\footnote{Since their own beliefs are included.}. This result is general, as formalized in the following proposition:  

\begin{proposition}[Vanishing of Posterior Variance]
    \label{prop:van_posterior}
    Assuming that the variance associated with each source of information is positive and bounded, then:  
    \begin{equation}
        \lim_{K \to \infty} \Tilde{\mathbb{V}}_t(\theta_{t+1}) = 0.
    \end{equation}
\end{proposition}  

\textit{Proof.} See Appendix \ref{sec:proof_van}.  

This proposition implies that uninformed agents become increasingly confident in their posterior beliefs as they receive more signals, eventually acting as if their expectations are certain. This feature results in a higher quantity demanded in the market compared to the baseline case, leading to increased volume and more volatile returns. However, the symmetry in the updating process ensures that the return distribution remains symmetric.  

The posterior mean for uninformed agents is given by:  
\begin{equation}
     \Tilde{\mathbb{E}}_t(\theta_{t+1}) = \lambda \theta_{t+1} + \xi \gamma_{t+1} + (1 - \lambda - \xi) \Tilde{\mathbb{E}}_{t-1}(\theta_{t}).
\end{equation}  

The resulting pricing equation is then:  
\begin{equation}
\begin{gathered}
    p_t = \frac{d}{r} + \left( \lambda \frac{ \frac{\theta_{t+1}}{R-\beta}}{V_I} + \xi \frac{ \frac{\gamma_{t+1}}{R-\beta}}{V_M} + (1 - \lambda - \xi) \frac{\frac{\Tilde{\mathbb{E}}_t(\theta_{t+1})}{R-\beta}}{V_U }\right) \\
    \left( \frac{\lambda}{V_I} + \frac{\xi}{V_M} + \frac{(1 - \lambda - \xi)}{V_U }\right)^{-1},
\end{gathered}
\end{equation}  

with:  
\[
V_I :=  \sigma^2_{\varepsilon} +
   \frac{\sigma^2_{\eta}}{(R-\beta)^2}, \quad V_M := \sigma^2_{\varepsilon}
   + \frac{\sigma^2_{\nu}}{(R-\beta)^2}, \quad V_U  := \sigma^2_{\varepsilon} + \frac{\sigma^2_{\eta}}{I}.
\]  

Rearranging, this reads as:  
\begin{equation}
    \begin{gathered}
    p_t = \frac{d}{r} + \frac{1}{R - \beta} \frac{\lambda \left( \theta_{t+1}\right) V_M V_U + \xi (\gamma_{t+1}) V_I V_U + (1 - \lambda - \xi) \Tilde{\mathbb{E}}_t(\theta_{t+1}) V_I V_M}{\lambda V_M V_U + \xi V_I V_U + (1 - \lambda - \xi)V_I V_M}.
    \end{gathered}    
\end{equation}  

\begin{figure}[H]
    \centering
    \includegraphics[width=0.6\textwidth]{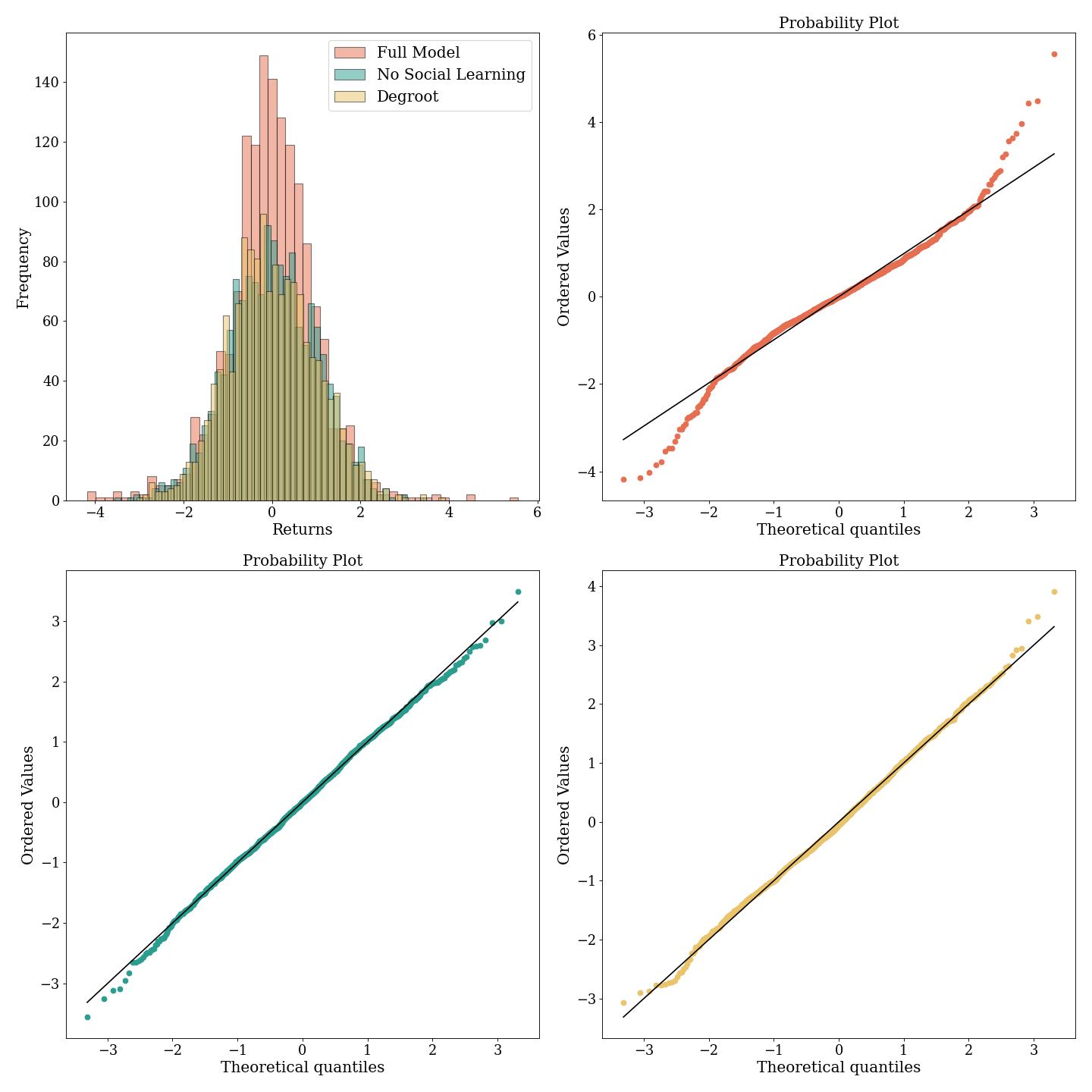}
    \caption{Return distributions for the three models.}
    \label{fig:model_comparison}
\end{figure}

We now compare the two models to the full model, in which agents are connected in a small-world network. To ensure comparability, we simulate the three models for 30 different realizations of the stochastic processes using different seeds and show the distribution of the normalized model-implied returns in Figure \ref{fig:model_comparison}. The parameters used are the same for all models and are fixed to the values in Table \ref{tab:param_calib}.  

We compare the return distributions for all models and plot a quantile-quantile (QQ) plot for each return series. Both the baseline and the DeGroot model exhibit symmetric return distributions and no excess kurtosis. The introduction of our social learning mechanism is not only more realistic but also necessary to match empirical evidence.  

In the next section, we turn to the calibration of the full model and the exploration of its properties.




\section{Empirical Calibration}\label{sec:calibration}
First, we summarize the model by offering a more accesible visualization of the sequence of events taking place in each time step $t$
\begin{enumerate}
    \item Agents update their prior beliefs regarding the observable component of dividends $\theta_{t+1}$, according to their own category;
    \item Agents compute the average forecast error over payoffs of the nodes they are connected to. The last payoff considered by them in the computation is $y_{t-1}$;
    \item Agents use the updating mechanism to obtain their posterior beliefs regarding the observable component of dividends $\theta_{t+1}$;
    \item The implied expected payoff $y_{t+1}$ is computed by each agent;
    \item The resulting price $p_t$ and individual demands are computed.
\end{enumerate}

Then we discipline the model and its many parameters by calibrating it with the following strategy.
We target the daily returns of the TESLA stock from September 2023 to September 2024.
In the literature a common choice to obtain a reliable calibration is to use some index,
typically the S\&P 500 index \citep{Schmitt2017}. In this context we argue that it is more natural to work with a single stock and in particular with one which features 
a high percentage of retail investors participation. 
In TESLA case the ownerhisp attributed to the public is around 40\%. 
Moreover individuals and insiders make up 12\% of the total ownership, 
thus making the stock an ideal candiate to study the effects of information frictions.
We then deal with the calibration of the model by following a three-step procedure.

First, some parameters are specific to the network topology. 
The one we use as our benchmark is the \cite{Watts1998} Small World Network. 
The network is generated by starting with a regular lattice of a given degree, assigning each node to one of the three categories introduced in Section \ref{sec:information_diffusion} according to specified proportions. Finally we rewire a fraction of edges randomly 
by a given probability of rewiring, introducing long-distance connections in the network. The main features of such a topology are local 
clustering, short-average path length and almost homogeneous degree of connection across nodes. 

Second we keep some parameters constant 
at values which are arbitrarily chosen after exprimentation with the model, as they do not affect the model dynamics. 
Total time steps are sufficient to ensure that behaviors driven by initial conditions are absorbed in the long run. 
The number of agents in the model is of relative importance only when combined with network specific parameters. 
The gross risk free rate is the daily equivalent corresponding to the average risk free rate in the period September 2023-September 2024 as proxied 
by the Market Yeald on U.S. Treasury Securities at 1-year Constant Maturity. 
The constant component of dividends\footnote{Although TESLA does not currently pay dividends, one could interpret this as an expected average dividend based on expected earnings and an expected payout ratio.
Mathematically this value has the only effect of shifting prices and has no impact on the dynamics.} is calibrated to match the daily closing price at the beginning of the period:
$d = p_0 \cdot r$.
The coefficient of constant risk aversion is at a value common in the literature, see for example \cite{Chetty2006}.

Third, for some parameters we explore their effect on the model for a wide range of values by Sobol sensitivity analysis 
\cite{Sobol2001}, \cite{Saltelli2002}, \cite{Saltelli2010} implemented by using the the SALib python library \citep{Herman2017}. 
Sobol sensitivity analysis is a global variance-based method used to quantify the contribution of individual input variables, 
or in our case parameters, to the output variance of a model. 
Given a model of the form \( Y = f(X_1, X_2, \dots, X_k) \), where \( Y \) is a scalar output, the first-order effect of a factor \( X_i \) can be expressed as

\begin{equation}
V_{X_i}\left( \mathbb{E}_{X_{\sim i}}(Y | X_i) \right),
\end{equation}
 
where \( X_i \) represents the \( i \)-th factor, and \( X_{\sim i} \) denotes the matrix of all factors except \( X_i \). The inner expectation operator \( \mathbb{E}_{X_{\sim i}}(Y | X_i) \) represents the mean of \( Y \) over all possible values of \( X_{\sim i} \) while keeping \( X_i \) fixed. The outer variance operator is taken over all possible values of \( X_i \). The associated sensitivity measure, known as the first-order sensitivity index, is then given by:

\begin{equation}
S_i = \frac{V_{X_i} \left( \mathbb{E}_{X_{\sim i}}(Y | X_i) \right)}{V(Y)}.
\end{equation}

\( S_i \) is normalized, as \( V_{X_i} \left( \mathbb{E}_{X_{\sim i}}(Y | X_i) \right) \) ranges between zero and \( V(Y) \). \( V_{X_i} \left( \mathbb{E}_{X_{\sim i}}(Y | X_i) \right) \) measures the first-order (additive) effect of \( X_i \) on the model output, while \( \mathbb{E}_{X_i}\left( V_{X_{\sim i}}(Y | X_i) \right) \) is customarily referred to as the residual.

Another commonly used variance-based sensitivity measure is the total effect index, defined as:

\begin{equation}
S_{T_i} = \frac{\mathbb{E}_{X_{\sim i}} \left( V_{X_i}(Y | X_{\sim i}) \right)}{V(Y)} = 1 - \frac{V_{X_{\sim i}} \left( \mathbb{E}_{X_i}(Y | X_{\sim i}) \right)}{V(Y)}.
\end{equation}

The total effect index \( S_{T_i} \) measures the total contribution of \( X_i \), which includes both the first-order and higher-order effects (interactions) of the factor. One way to interpret this is by considering that \( V_{X_{\sim i}} \left( \mathbb{E}_{X_i}(Y | X_{\sim i}) \right) \) represents the first-order effect of \( X_{\sim i} \), so that \( V(Y) \) minus \( V_{X_{\sim i}} \left( \mathbb{E}_{X_i}(Y | X_{\sim i}) \right) \) must give the contribution of all terms in the variance decomposition that include \( X_i \).

\begin{figure}[H]
    \centering
    \includegraphics[width = 0.9\textwidth]{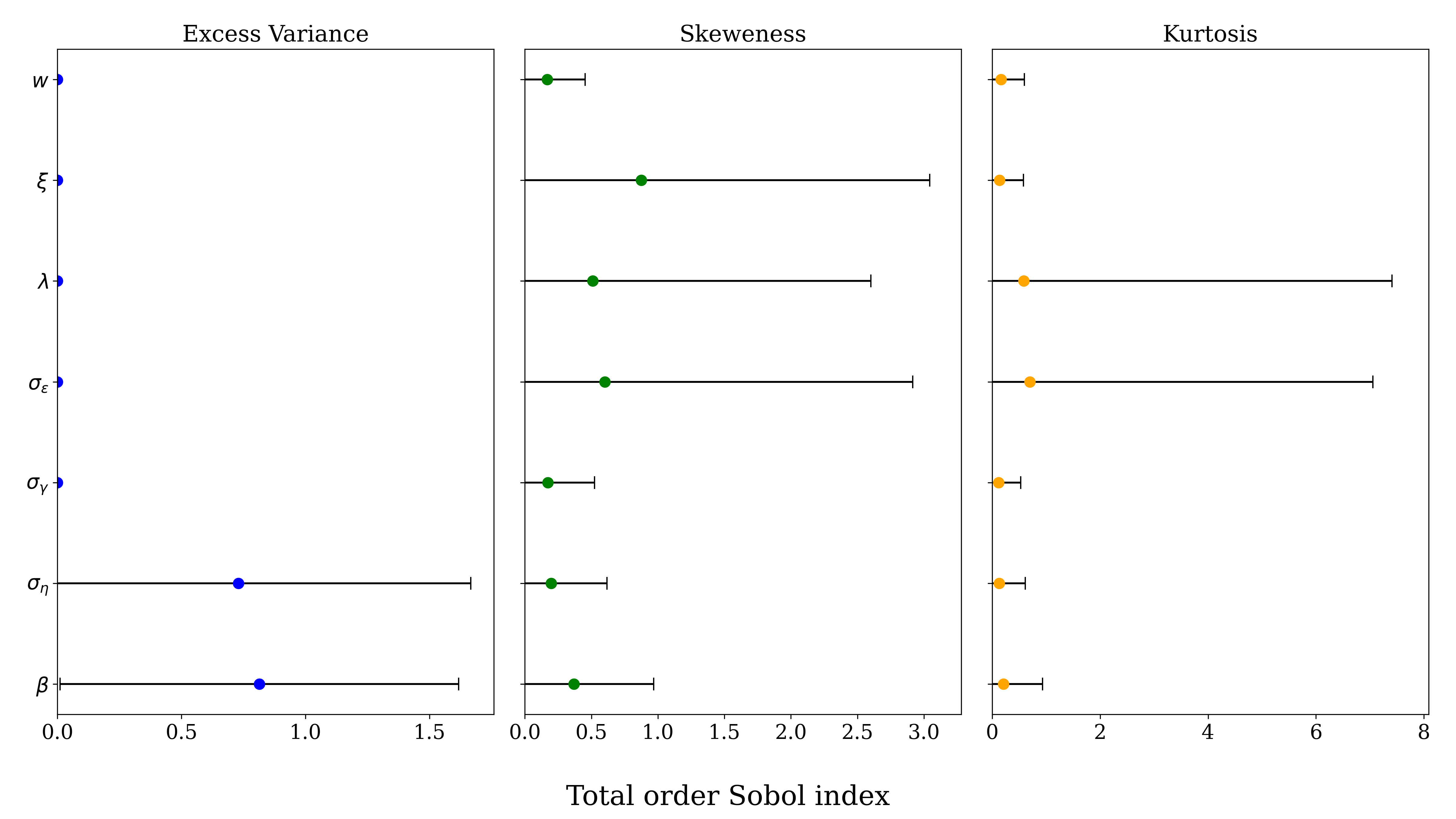}
    \caption{Sobol Sensititivty Analysis}
    \label{fig:sobols_sen}
\end{figure}

In figure \ref{fig:sobols_sen} we report the total order Sobol index for the seven parameters 
in the y-axis, computed on three outputs of the model, namely: price variance in excess of a model with a representative fully informed investor
which we use as benchmark, skeweness and kurtosis of the model returns.
All results are obtained from using 128 samples for each of the 9 parameters of interest, with 30 different seeds per combination, resulting in a total of 34560 simulations.

Excess variance seems to be driven by two parameters, $\beta$ and $\sigma_{\eta}$, which combine 
to determine the unconditional variance of the observable component of dividends.
The skeweness of the model returns is driven by the proportions of informed and misinformed agents, $\lambda$ and $\xi$,
and by the size of the unobservable noise component of dividends $\sigma_{\varepsilon}$.
A similar result is obtained for the kurtosis of the model returns, altough now the proportion of misinformed agents seems to play a smaller role.
The number of informed and misinformed agents in the model however has both a direct and indirect effect.
The former is due to the fact that these agents are also actively operating in the market and therefore their beliefs are directly reflected in the price.
The latter is due to the effect that their communication has on the beliefs of uninformed agents, which in turn affect the price.
Since we are mainly interested in this last effect we decide to keep the proportion of informed and misinformed agents at a constant value.
The value we chose is 5\% of the population for both classed, in order to have a reasonable number of agents in each category while still being able
to attribute the majority of the price dynamics to the uninformed agents.
The volatilities of the unobservable and observable component of dividends combine to determine the noise-to-signal ratio
of the information.
We therefore normalize the value of $\sigma_{\varepsilon}$ to 1.
Finally, as already mentioned, $\beta$ and $\sigma_{\eta}$ jointly determines the variance of the information.
We therefore fix one of them, namely $\beta = 0.5$ and use the other to vary the variance of the information.
This leaves us with two parameters to calibrate $\sigma_{\eta}$ and $\sigma_{\nu}$ 
and we do so by using two moments of the TESLA stock returns: skewness and kurtosis.
For this we use the Sequential Neural Posterior Estimation (SNPE) proposed by \cite{NIPS2016_6aca9700} and popularised to economics ABMs
by the recent work of \cite{DYER2024104827}.
SNPE is a likelihood-free inference method that uses neural networks to directly approximate the posterior distribution of model parameters.\footnote{This is in contrast with classical Bayesian methods in which the posterior is computed as the product of the likelihood and the prior.} 
The key idea behind this method is to simulate data from the model at different parameter values and use a neural network to learn the conditional density \( q_{\phi}(\theta | x) \), where \( \theta \) are the parameters and \( x \) is the observed data (e.g., TESLA stock skewness and kurtosis).

The process works as follows:
\begin{enumerate}
    \item \textbf{Simulate Data}: For a given set of parameters \( \theta_n \), simulate data \( x_n \) from the ABM.
    \item \textbf{Neural Network Training}: Use the simulated parameter-data pairs \( (\theta_n, x_n) \) to train a Mixture Density Network (MDN). 
    The MDN outputs a conditional probability distribution \( q_{\phi}(\theta | x) \), which serves as an approximation to the true posterior \( p(\theta | x) \).
    \item \textbf{Posterior Approximation}: After training, the neural network provides an approximation of the posterior distribution for any observed data \( x_o \), by maximizing the likelihood
    \begin{equation}
        \max_{\phi} \frac{1}{N} \sum_{n=1}^{N} \log q_{\phi}(\theta_n | x_n).
    \end{equation}
    \item \textbf{Sequential Refinement}: The method iteratively refines the prior distribution \( p(\theta) \) based on the learned posterior, improving the efficiency of the simulations by focusing on plausible regions of the parameter space.
\end{enumerate} 

We refer the reader to \cite{DYER2024104827} for an in depth evaluation of the method.
While other methods such the Simulated method of moments as in \cite{Franke2012} and \cite{Franke2016} could be used, this method is particularly efficient
as it requires only enough simulations to train the neural network.
Moreover we assess the ability of the method to recover the true posterior by using a synthetic dataset generated by the model
in appendix \ref{sec:SNPE_validation}, which serves to show that in our setting the neural network approximation is accurate.

\begin{figure}
    \centering
    \includegraphics[width = 0.6\textwidth]{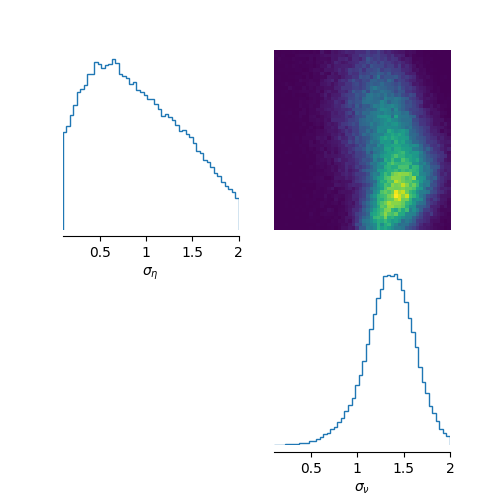}
    \caption{Posterior Distribution}
    \label{fig:posterior}
\end{figure}

For the real data estimation we take an agnostic position on the prior. 
Given that we are estimating variances we use a uniform prior distribution, with support $[0.1, 2]$ and report 
the results as well as the other parameter values in table \ref{tab:param_calib}. 
The posterior distribution is extremely well behaved and uni-modal as we can report in Figure \ref{fig:posterior}, with median values
$\sigma_{\eta} = 0.84$ and $\sigma_{\nu} = 1.36$.
It is interesing to notice that shocks to misinformation are slighlty lager than shocks to information.
This is consistent with misinformation being more sensational than actual information, in order to generate more attention.

\begin{table}[H]
    \caption{Parameter Calibration}
    \label{tab:param_calib}
     \begin{adjustbox}{width=\textwidth}
    \begin{tabular}{cccccc}
    \hline \hline
    \multicolumn{6}{c}{\textbf{Calibrated Parameters}}                                                                                                          \\ \hline
    Parameter              & Range Explored    & Value  & \multicolumn{2}{c}{Source}                       & Description                                        \\ \hline
    T                      & -                 & 1500   & \multicolumn{2}{c}{Own Calibration}                            & Total Time Steps                                   \\
    I                      & -                 & 200    & \multicolumn{2}{c}{Own Calibration}                            & Number of Agents                                   \\
    R                      & -                 & 1,0001 & \multicolumn{2}{c}{Average daily risk free rate 09:2023 - 09:2024}                             & Gross Risk Free Rate                               \\
    d                      & -                 & 0,021  & \multicolumn{2}{c}{Directly calibrated}           & Constant Component of Dividends                    \\
    a                      & -                 & 1    & \multicolumn{2}{c}{\citep{Chetty2006}}                             & Coefficient of constant risk aversion              \\
    $\sigma_{\varepsilon}$ & ${[}0.0001, 2{]}$ & 1      & \multicolumn{2}{c}{Normalized}                   & Std of the Unobservable Component of Dividends     \\
    $w$                    & $[0.1, 0.9]$      & 0.9    & \multicolumn{2}{c}{Own Calibration}                      & Memory parameter of the Exponential Moving Average \\
    $\lambda$              & $[0.01, 0.3]$     &    0.05    & \multicolumn{2}{c}{Own Calibration} & Proportion of Informed Agents                      \\
    $\xi$                  & $[0.01, 0.3]$     &     0.05   & \multicolumn{2}{c}{Own Calibration}                             & Proportion of Misinformed Agents                   \\ \hline \hline
    \multicolumn{6}{c}{\textbf{Estimated Parameters}}                                                                                                           \\ \hline
    Parameter              & Prior             & \multicolumn{3}{c}{Posterior}                             & Description                                        \\ \hline
                           &                   & Mean   & Median                   & Std                   &                                                    \\ \cline{3-5}
    $\sigma_{\eta}$      & Uniform (0,2)     &   0.89     &  0.84                        &           0.47            & Std of Information shocks     \\
    $\sigma_{\nu}$      & Uniform (0,2)     &    1.34    &     1.36                     &            0.26           & Std of Misinformation shocks               \\ \hline \hline
    \end{tabular}
    \end{adjustbox}
    \end{table}

\section{Numerical Simulations}\label{sec:numerical_simulations}
\begin{figure}[H]
    \centering
    \begin{subfigure}[b]{0.32\textwidth}
        \centering
        \includegraphics[width=\textwidth]{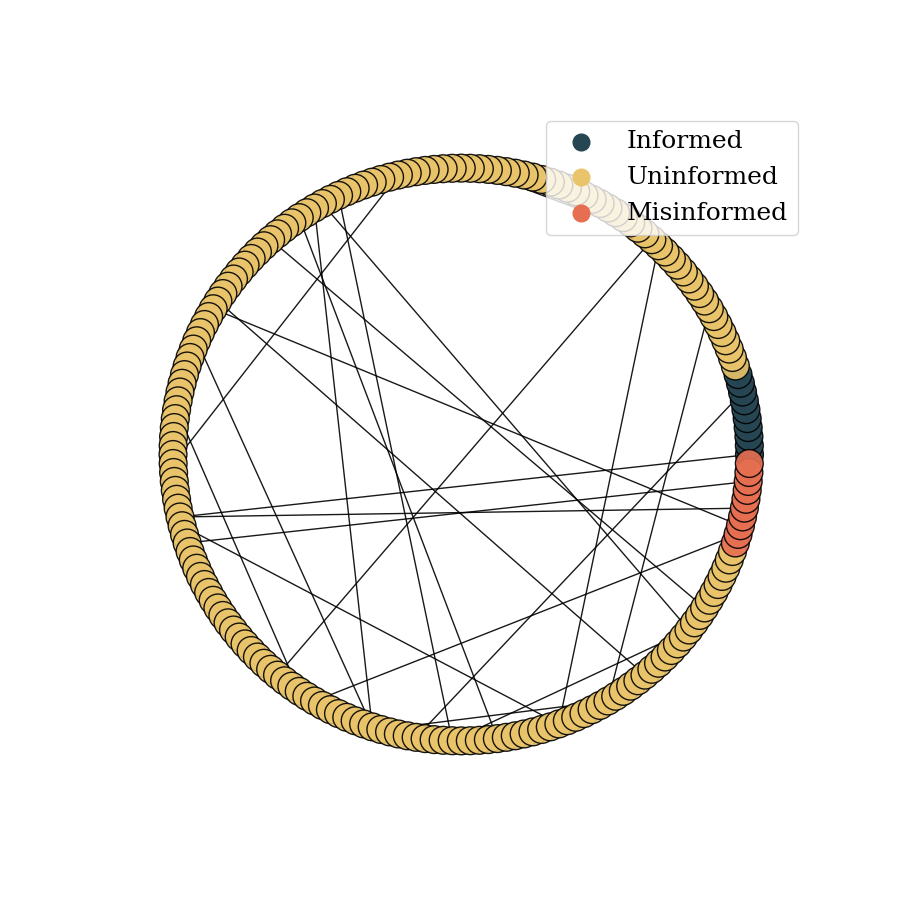} 
        \caption{Network Structure}
        \label{fig:imagea}
    \end{subfigure}
    \begin{subfigure}[b]{0.32\textwidth}
        \centering
        \includegraphics[width=\textwidth]{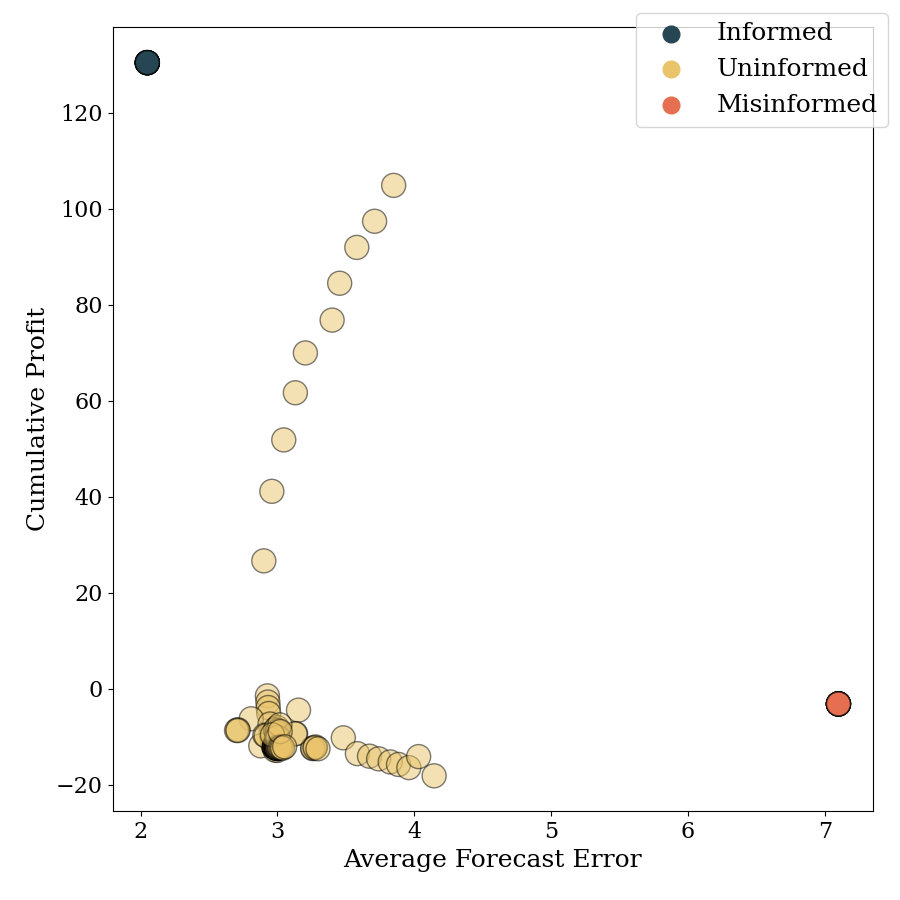} 
        \caption{Cumulative Profits}
        \label{fig:imageb}
    \end{subfigure}
    \begin{subfigure}[b]{0.32\textwidth}
        \centering
        \includegraphics[width=\textwidth]{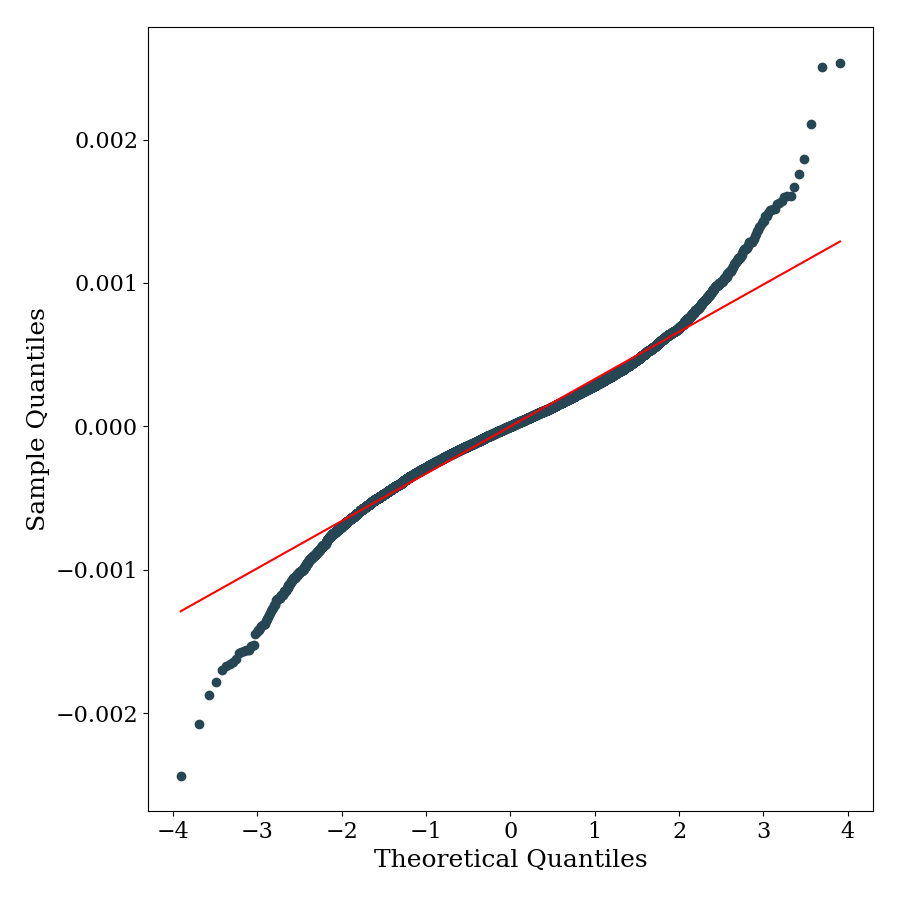} 
        \caption{QQ Plot}
        \label{fig:imagee}
    \end{subfigure}
    \caption{Small World Network}
    \label{fig:small_world}
    \caption*{\footnotesize{Note: Data are obtained from 30 simulations with different stochastic seeds. 
    \textbf{Network parameters} are: Network density = 0.1, Probability of rewiring = 0.01.}}
\end{figure}

With our calibration we can then show key features of the model.
We do so in Figure \ref{fig:small_world} by using the average over 
30 Montecarlo simulations with different stochastic seeds. In panel (\ref{sub@fig:imagea}) we plot the network structure used in the simulations. 
Agents position is randomly determined in the beginning on the experiment and then kept fixed, in order to allow us to track each 
agent evolution in the different simulations. 
Panel (\ref{sub@fig:imageb}) shows a scatter plot of cumulative profits and average forecast error 
for each agent.The accuracy of uninformed agents lies between that of the other two categories. In this configuration, misinformation does not appear to spread significantly into the network. Although there is a generally linear positive relationship between accuracy and profits, there are some notable exceptions. While most uninformed investors incur small gains or even losses, a small subset earns large profits, sometimes almost as high as those of informed agents.  
The reason for this pattern is that these agents are directly linked to a source of information. They receive information relatively early and act on before others, allowing them to profit the most from the delayed price adjustment caused by the gradual spread of information in the network.
Moreover as already pointed out, uninformed agents with many connections might have a lower posterior variance than informed agents, therefore taking larger positions
and profiting more from the price movement in the direction of their beliefs.
Lastly we look at the impact on returns. 
In panel (\ref{sub@fig:imagee}) we use a QQ plot to identify the presence of fat tails. We compare the simulated quantiles 
with the theoretical quantiles from a Normal distribution with the same mean and standard deviation of model's returns. 
In table (\ref{tab:summary}) we report the summary statistics of the model returns. 
Skeweness and kurtosis are the target moments we have used for the calibration and are 0.04 and 5.50 in the simulated data and 0.25\footnote{While normally one would expect negative skeweness in financial markets, the 
period we considered was associated with a series of favorable earnings reports and positive growth expectations.} and 5.54 in the empricial counterpart.
We also report the profit (or loss) incurred by the misinformed agents in the model. In this case it is negative, showcasing that while misinfomration is spreading in the netowork
it does not diffuse outside of the local cluster of connections to misinformed agents.

We can then analyze the implications of the model for the absorption of information in the market using an impulse response function (IRF) analysis. 
In our framework, shocks propagate through two distinct channels.  
First, there is a \textit{direct channel} as the shock directly influences expectations, affecting investors’ demands and ultimately feeding into market prices. 
Second, there is an \textit{indirect channel} as the resulting prices and communication between agents feed back into how uninformed traders update their weighting matrices, 
which determine the relative importance they assign to different sources of information.  

\begin{figure}[H]
    \centering
    \includegraphics[width = 0.9\textwidth]{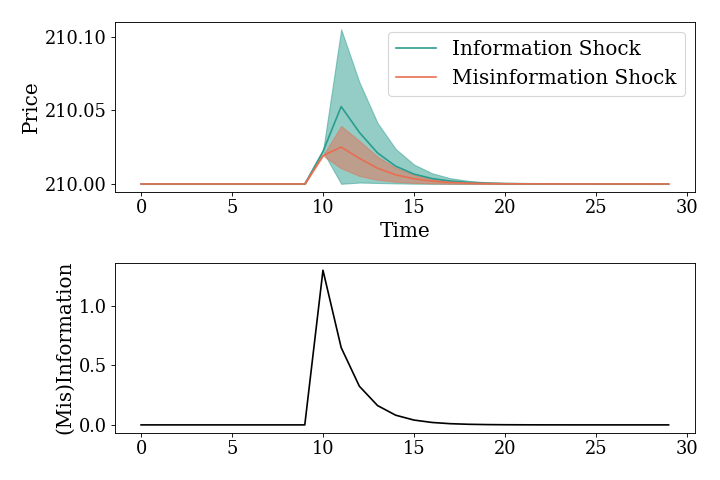}
    \caption{Impulse Response To a (Mis)Information Shock at time $\tau = 10$}
    \label{fig:impulse_response}
\end{figure}

If we were to introduce a shock to the \(\theta\) process at time \(\tau\), we could not attribute the resulting effects only to this specific shock because earlier shocks would still influence both the network’s structure and the ongoing diffusion of information.  
To isolate the causal impact of a single shock, we proceed as follows. 
We begin by simulating the model for a sufficiently long period to ensure that the weighting matrix converges to a stable representative state. 
We then record this matrix and use it to initialize a new market scenario in which no further updates to the weighting matrix occur. 
By suppressing all noise until time \(\tau = 10\), agents have no reason to alter their weightings, which therefore remain fixed at the initial distribution.  
Under these controlled conditions, we introduce a one-standard-deviation shock to either \(\theta\) (an information shock) or \(\gamma\) (a misinformation shock). 
This setup allows us to compute the resulting impulse response and entirely capture the direct effect of the shock. 
Figure \ref{fig:impulse_response} shows the impulse response of the price to a one-standard-deviation shock.  

The results align with patterns observed in empirical studies. 
While misinformation shocks do propagate into market prices, their overall impact is weaker than that of information shocks. 
In line with the findings of \cite{Clarke2021}, the market appears able to filter out some of the noise generated by misinformation.  
By contrast, information shocks not only exert a stronger influence on prices but also continue to shape price dynamics over time. 
In both cases, the delayed peak of the IRF is consistent with the assumption that information diffuses gradually through the network and only reaches its full impact on prices once uninformed agents have processed it \citep{Huberman2001}.

\subsection{The effect of different Network Topologies}\label{sec:network_topologies}
The use of social networks has drastically reshaped communication in recent years. The network topology plays a crucial role in enabling rapid information sharing but also raises concerns about how beliefs are formed.
In recent years, social media platforms have shifted from a friend-based structure, where connections are reciprocal, to a follower-based structure, where a few influencer nodes can have a disproportionately large number of connections. Moreover, the way many algorithms are designed to maximize user engagement can lead to the creation of echo chambers, where individuals are exposed only to information that confirms their existing beliefs.
While the small-world network structure we have used so far is a good approximation of the friend-based structure, we now explore the impact of polarization and follower-based structures on the model dynamics by adopting different network topologies. Crucially, to isolate the effect of the network structure, we repeat the analysis while keeping the variable parameters fixed.

\vspace{5mm}

\noindent
\textbf{Stochastic Block Network}

\begin{figure}[H]
    \centering
    \begin{subfigure}[b]{0.32\textwidth}
        \centering
        \includegraphics[width=\textwidth]{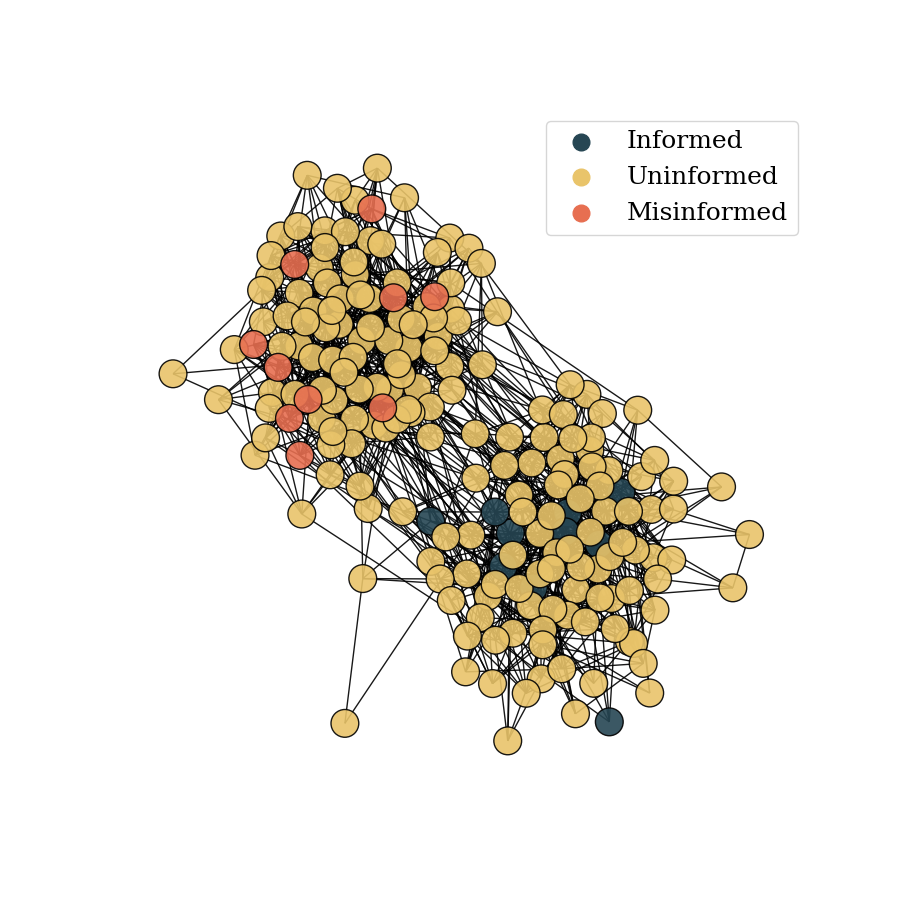} 
        \caption{Network Structure}
        \label{fig:image_sbm_a}
    \end{subfigure}
    \begin{subfigure}[b]{0.32\textwidth}
        \centering
        \includegraphics[width=\textwidth]{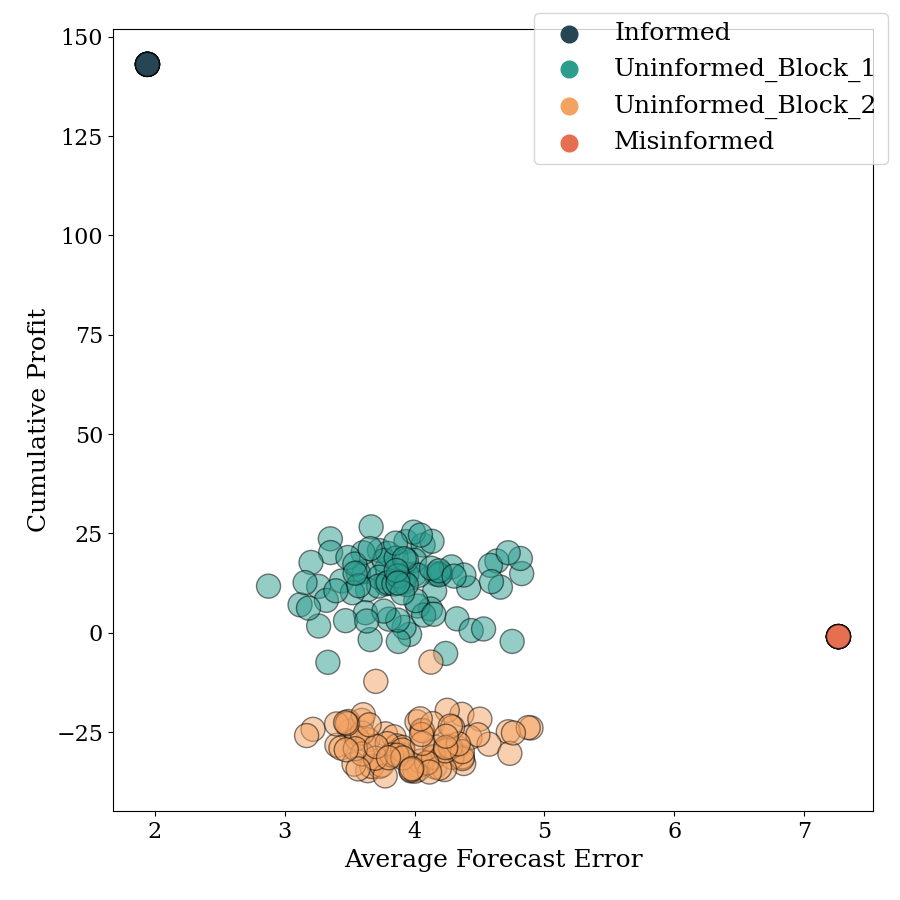} 
        \caption{Cumulative Profits}
        \label{fig:image_sbm_b}
    \end{subfigure}
    \begin{subfigure}[b]{0.32\textwidth}
        \centering
        \includegraphics[width=\textwidth]{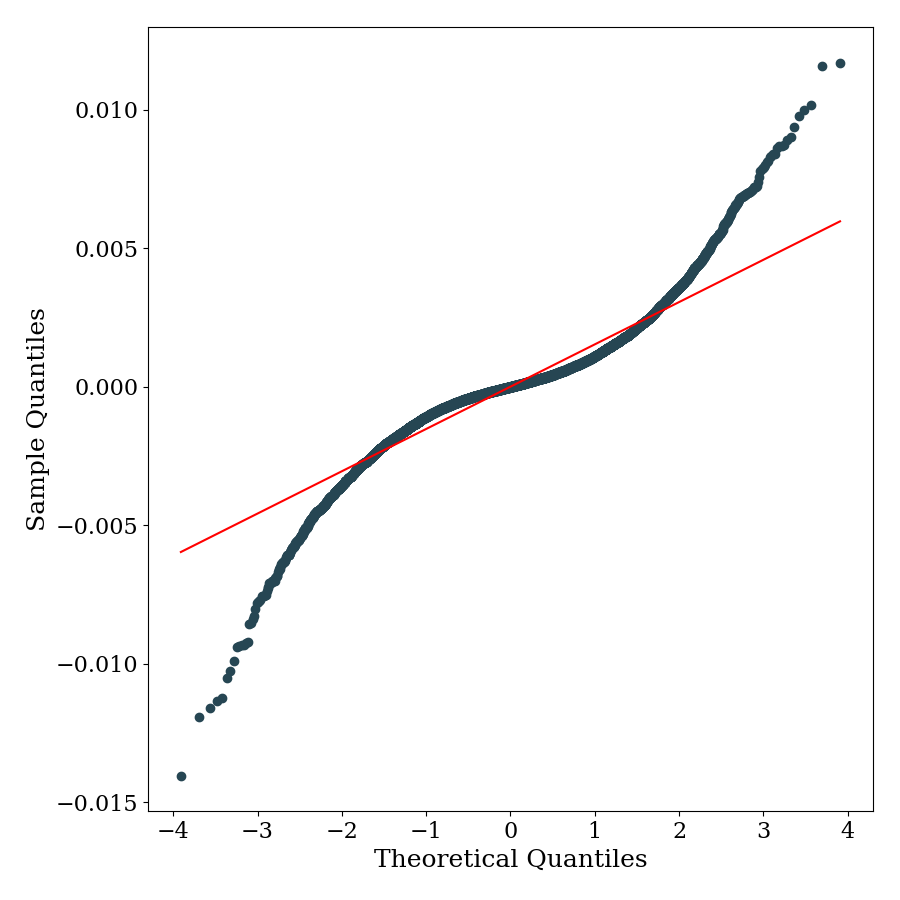} 
        \caption{QQ Plot}
        \label{fig:image_sbm_d}
    \end{subfigure}

    \caption{Stochastic Block Network}
    \label{fig:price_SBM}
    \caption*{\footnotesize{Note: Data are obtained from 30 simulations with different stochastic seeds. 
    \textbf{Network parameters} are: Number of Partitions = 2, Density of Intra-Groups Edges = 0.1, 
    Density of Inter-Groups Edges = 0.001.}}
\end{figure}

The first scenario we analyze is a completely polarized society. We create it by using a Stochastic Block Model \cite{Holland1983},
in which we partition the nodes in order to create two clusters. 
Informed and misinformed agents are separated and assigned to either the information block or the misinformation one. 
We denote by density of intra-groups edges, the likelihood of having an edge between agents belonging to the same cluster. 
This is higher than the density of inter-groups edges, regulating connections between agents belonging to different blocks. 
Thus not only the network is partitioned, but communication between the two groups is scarce. 
We report the network specific parameters in table (\ref{tab:SBM_parameters}).
\begin{table}[H]
\centering
\caption{Parameters of the Stochastic Block Network}
\label{tab:SBM_parameters}
\begin{tabular}{lc}
\hline \hline
\textbf{Parameter}               & \textbf{Value} \\ \hline
Number of Partitions                  & $2$  \\
Density of Intra-Groups Edges          & $0.1$   \\ 
Density of Inter-Groups Edges          & $0.001$ \\ \hline \hline
\end{tabular}
\end{table}
As before we simulate the model 30 times with different stochastic seeds. Results are displayed in Figure (\ref{fig:price_SBM}).

Panel (\ref{sub@fig:image_sbm_a}) illustrates the network structure, which now consists of two clusters each exhibiting small-world architectures. 
Under these conditions, uninformed agents are less likely of receiving reliable information and misinformation simultaneously. 
The consequences of this segregation are clearly visible in Panel (\ref{sub@fig:image_sbm_b}), where agents in the information block earn higher profits than those in the misinformation block.
Panel (\ref{sub@fig:image_sbm_d}) shows that, although the overall shape of the returns distribution remains qualitatively similar, the leptokurtosis is now more pronounced. 
Misinformed agents continue to incur losses, though these losses are smaller than in the baseline scenario. 
While these agents can influence their own cluster, their erroneous beliefs fail to fully penetrate the market’s pricing mechanism. 
As a result, uninformed agents situated in the informed cluster capitalize on their disadvantaged counterparts in the misinformed segment of the network.

\vspace{5mm}

\noindent
\textbf{Scale Free Network}
Until now we have simulated societies in which agents have equal opportunities of sharing their beliefs, 
given that the average degree was rather homogeneous across the network. 
For our next scenarios we opt to work with societies in which certain individuals can have a disproportionate impact on the network\footnote{There are multiple example of investment
platforms building on this feature. A famous one is the multi-asset investment plaform eToro, in whcih users can directly copy the porfolio allocation of \textit{top-performing investors}. }. 
This concept is similar to the one of ``guru", discussed in \cite{Tedeschi2012}. 
Gurus are are agents that are most imitated by others and can emerge endogenously in the market. 
The main difference is that in their model, edges are created by a mechanism of preferential attachment based on wealth. 
Instead we use an exogenous mechanism so by creating a directed\footnote{It must be remarked that although until now we were using 
undirected networks, informed and misinformed agents were behaving in a dogmatic fashion. 
This is because a $0$ prior variance in their beliefs implies non updating or posterior beliefs exactly equal to the prior.} 
Scale-Free Network \citep{Bollobas2003} and forcefully allocating either informed or misinformed agents in the nodes with 
most outward connections.
The network specific parameters are reported in table (\ref{tab:scale_free_parameters})
\begin{table}[H]
\centering
\caption{Parameters of the Scale Free Network}
\label{tab:scale_free_parameters}

\begin{tabular}{lc}
\hline \hline
\textbf{Parameter}               & \textbf{Range} \\ \hline
Alpha                 & $0.41$  \\
Beta         & $0.54$   \\ 
Gamma &  0.05 \\ \hline \hline
\end{tabular}
\caption*{\footnotesize{Note: Alpha is the probability for adding a new node connected to an existing node chosen randomly according to the in-degree distribution.
 Beta is the probability for adding an edge between two existing nodes.
 Gamma is the robability for adding a new node connected to an existing node chosen randomly according to the out-degree distribution.
 For a detailed explanation of the parameters role we refer to page 2 of \cite{Benabou1992}.}}
\end{table}

\begin{figure}[H]
    \centering
    \begin{subfigure}[b]{0.32\textwidth}
        \centering
        \includegraphics[width=\textwidth]{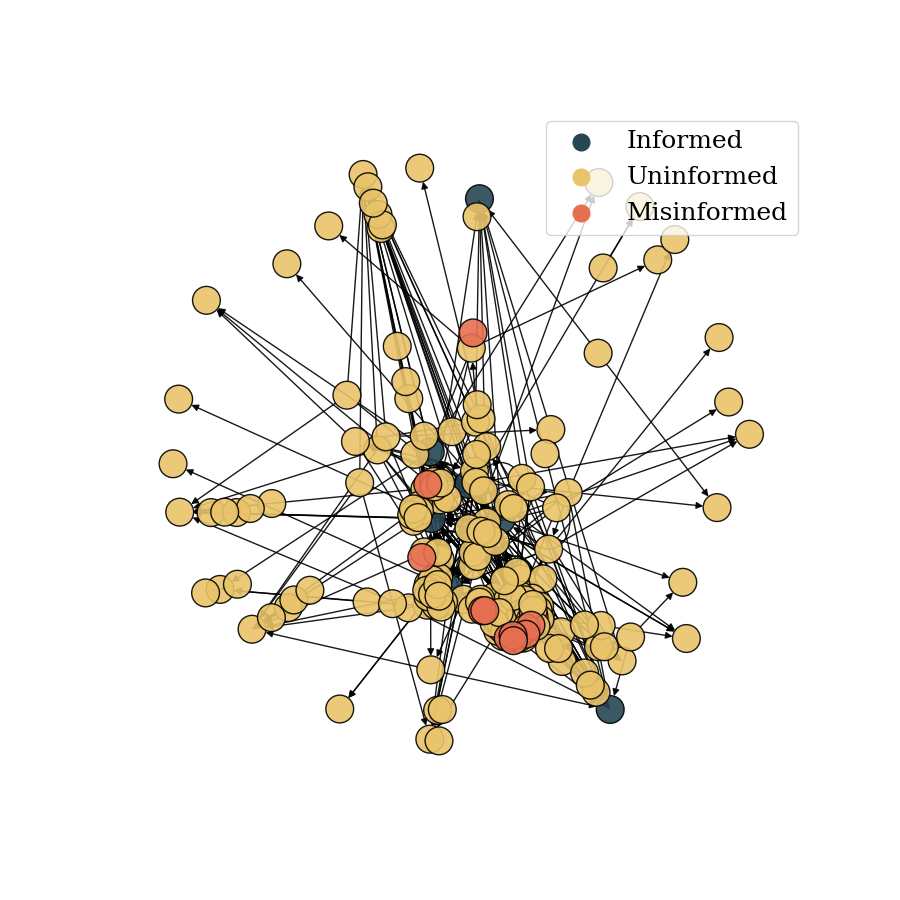} 
        \caption{Network Structure}
        \label{fig:image_sf_info_a}
    \end{subfigure}
    \begin{subfigure}[b]{0.32\textwidth}
        \centering
        \includegraphics[width=\textwidth]{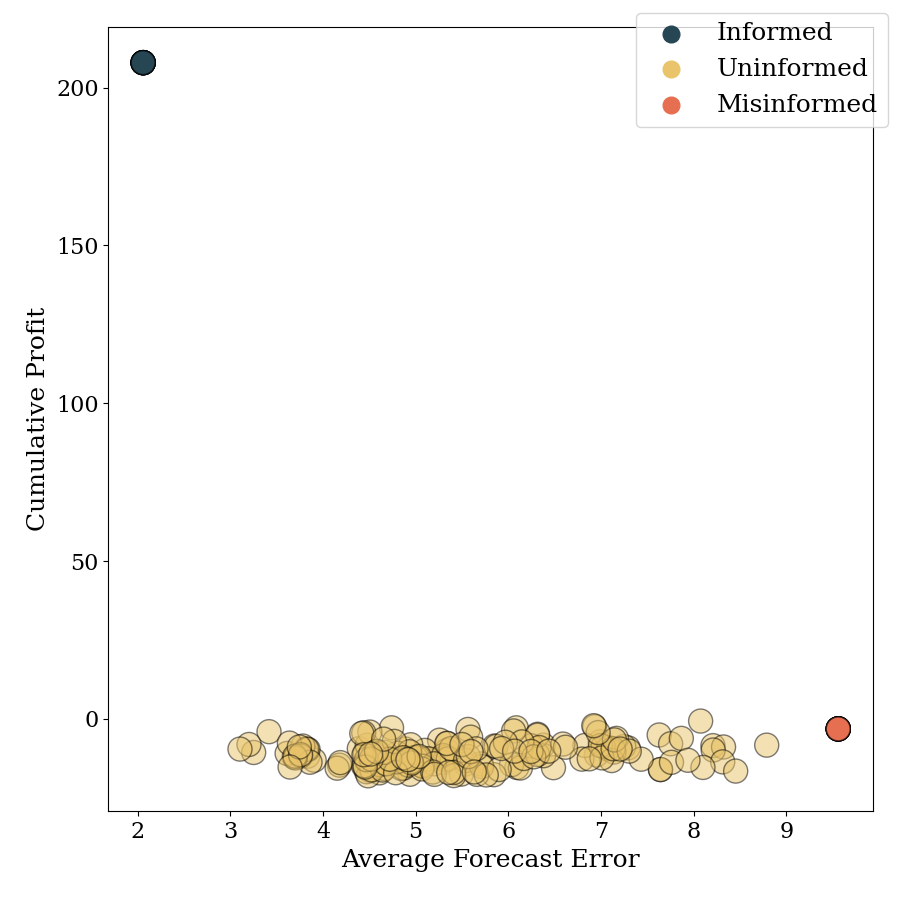} 
        \caption{Cumulative Profits}
        \label{fig:image_sf_info_b}
    \end{subfigure}
    \begin{subfigure}[b]{0.32\textwidth}
        \centering
        \includegraphics[width=\textwidth]{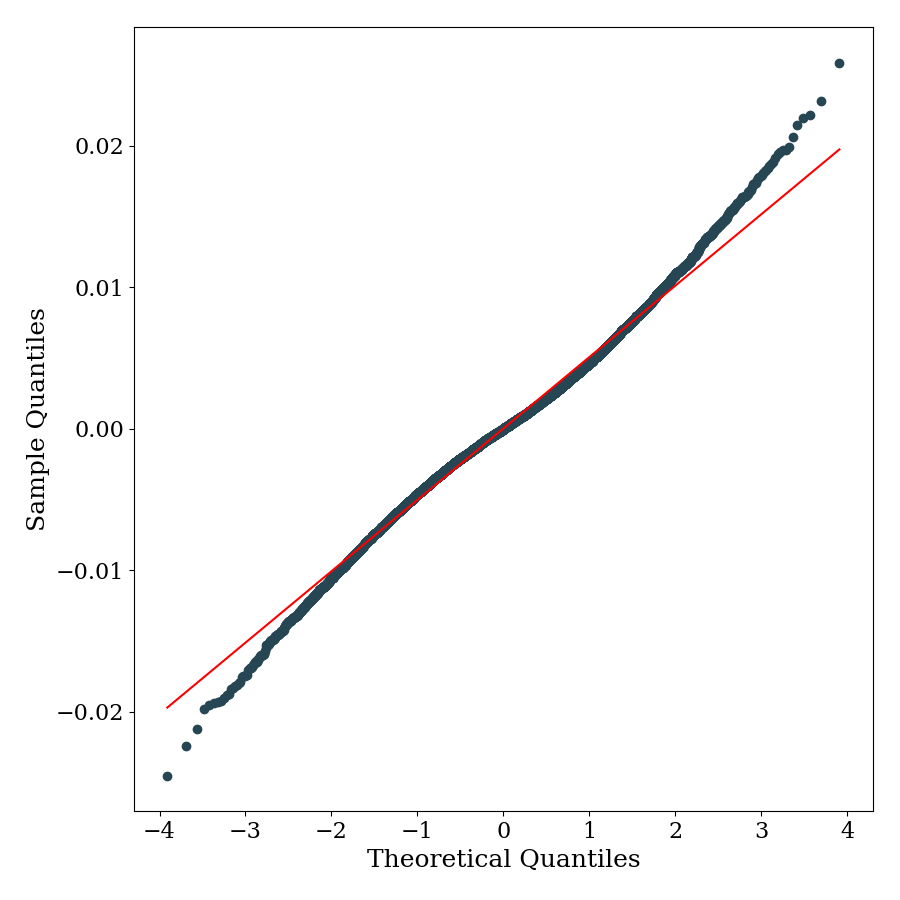} 
        \caption{QQ Plot}
        \label{fig:image_sf_info_d}
    \end{subfigure}
    \caption{Barabasi-Albert Graph, Informed}
    \label{fig:price_sf_info}
    \caption*{\footnotesize{Note: Data are obtained from 30 simulations with different stochastic seeds. 
    \textbf{Network parameters} are: Alpha = 0.41, Beta = 0.54, Gamma = 0.05.}}
\end{figure}

We begin by analyzing the case where informed agents are the most central. The network topology is displayed in panel (\ref{sub@fig:image_sf_info_a}) of Figure (\ref{fig:price_sf_info}). Panel (\ref{sub@fig:image_sf_info_b}) highlights that informed agents benefit significantly from their prominent position in the network, earning much more than in previous scenarios. This is because they can directly reach most of the uninformed traders.  
Since communication is lagged, informed agents push the price in the direction in which they have already taken a position. This results in the majority of uninformed agents incurring small losses. As everyone receives the information almost simultaneously, there are no additional gains to be made; in fact, uninformed agents lose money on average to informed agents who acted earlier based on their private information.  
Even so, it is still better for uninformed agents to incorporate stale news into their forecasts, given the persistence of these shocks. 
This result offers a potential explanation for the empirical findings regarding stale information reported by \cite{Gilbert2012} and \cite{Tetlock2011}.  
The effect of this configuration on returns is a distribution with extremely low kurtosis and skewness, as shown in table (\ref{tab:summary}) and panel (\ref{sub@fig:image_sf_info_d}). The market is as efficient as possible, given the inability of uninformed agents to access private information. As a result, there is almost no room for extremely large returns or extreme losses.

\begin{figure}[H]
    \centering
    \begin{subfigure}[b]{0.32\textwidth}
        \centering
        \includegraphics[width=\textwidth]{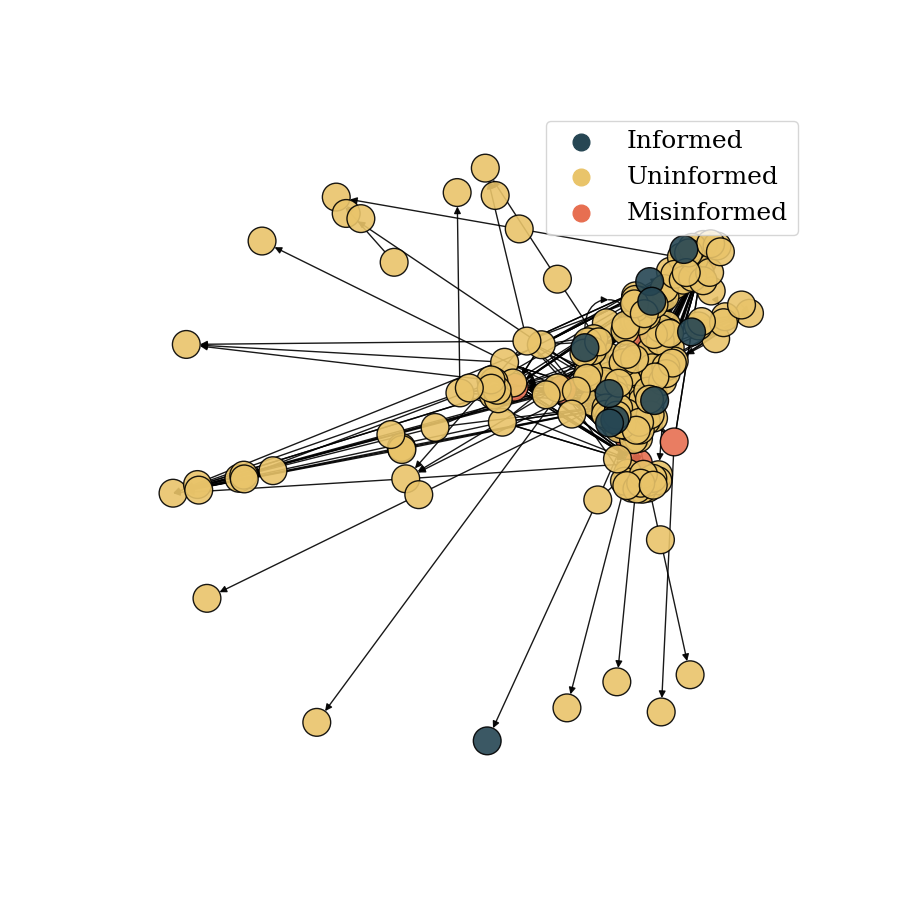} 
        \caption{Network Structure}
        \label{fig:image_sf_misinfo_a}
    \end{subfigure}
    \begin{subfigure}[b]{0.32\textwidth}
        \centering
        \includegraphics[width=\textwidth]{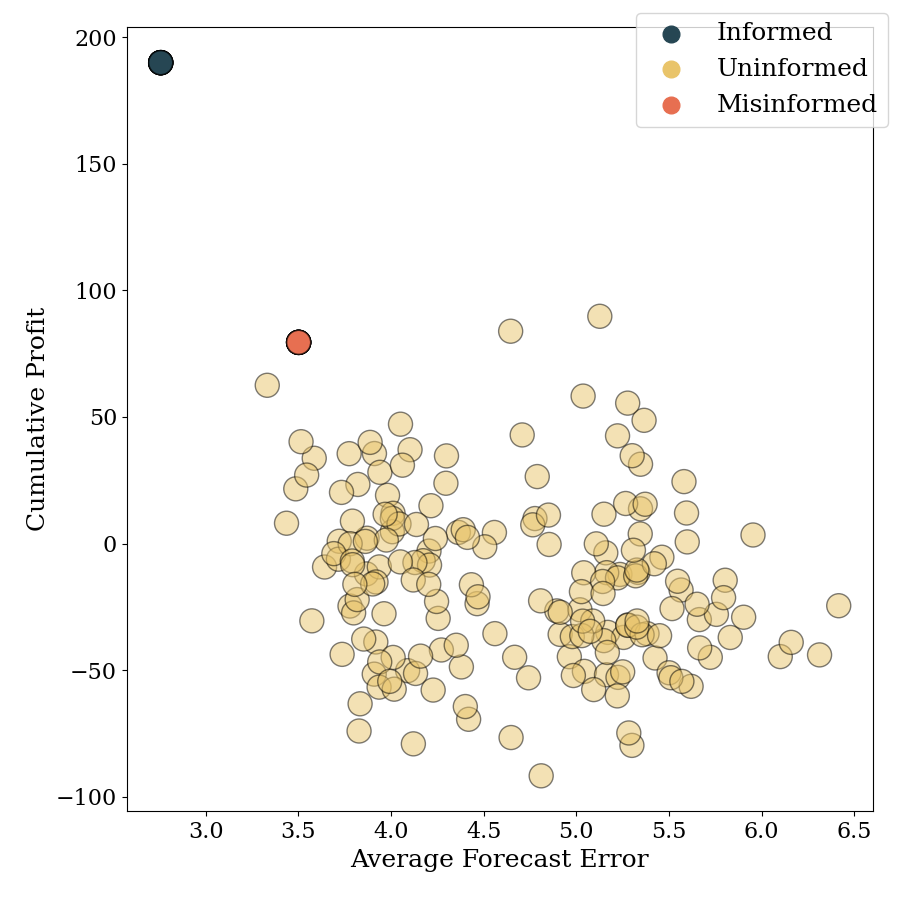} 
        \caption{Cumulative Profits}
        \label{fig:image_sf_misinfo_b}
    \end{subfigure}
    \begin{subfigure}[b]{0.32\textwidth}
        \centering
        \includegraphics[width=\textwidth]{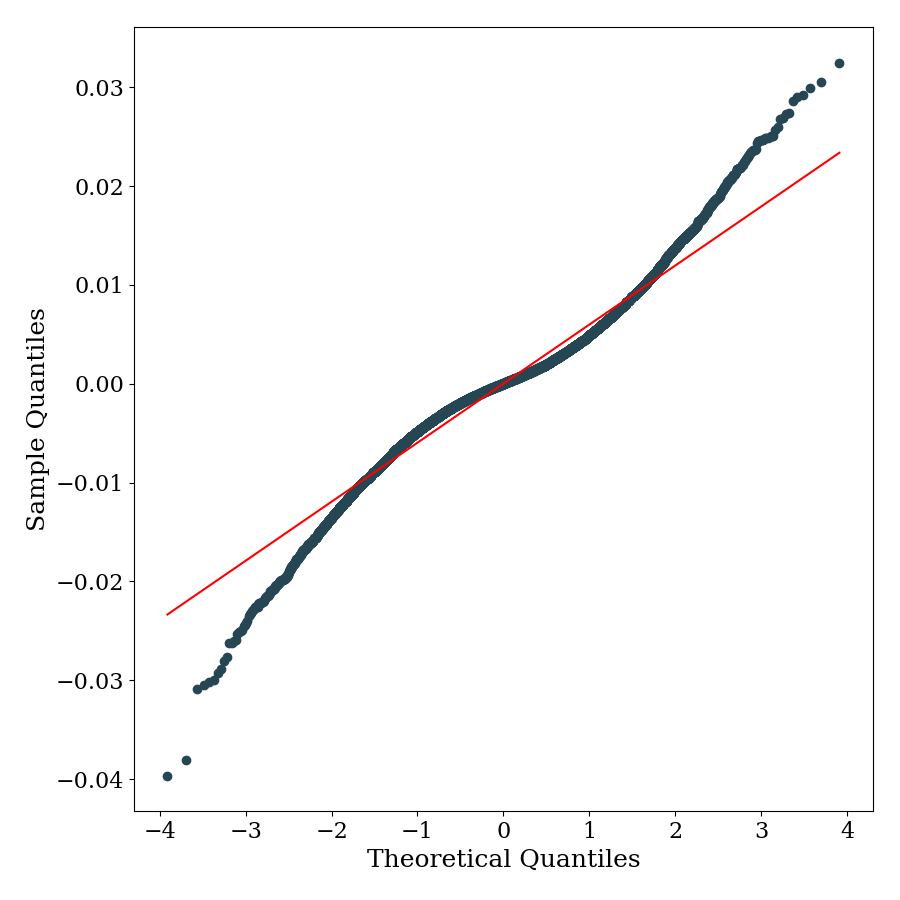} 
        \caption{QQ Plot}
        \label{fig:image_sf_misinfo_d}
    \end{subfigure}
    \caption{Barabasi-Albert Graph, Misnformed}
    \label{fig:price_sf_misinfo}
    \caption*{\footnotesize{Note: Data are obtained from 30 simulations with different stochastic seeds. 
    \textbf{Network parameters} are: Alpha = 0.41, Beta = 0.54, Gamma = 0.05.}}
\end{figure}

We then analyze the second scale-free society, in which the most connected nodes are misinformed agents. The network in panel (\ref{sub@fig:image_sbm_a}) of Figure (\ref{fig:price_sf_misinfo}) has the same configuration as the previous one, with the only difference being the position of agents. Turning to the profit and accuracy analysis in panel (\ref{sub@fig:image_sbm_b}), we can see that misinformation has successfully spread throughout the network. Misinformed agents now achieve positive and significantly high profits.  
This comes at the expense of uninformed agents, who are both losing more on average and showing a high degree of inequality in their profits. Having, in most cases, connections only to misinformed individuals or other uninformed agents with stale misinformation makes their participation in the market extremely unfruitful.  
Panel (\ref{sub@fig:image_sbm_d}) confirms that returns are again extremely leptokurtic, as reported in table (\ref{tab:summary}), and the skewness is now negative. This market is the least efficient of all the scenarios we have analyzed. However, some uninformed agents are still able to make profits. This is because the network structure allows certain agents to profit from future agents incorporating their beliefs, even though these beliefs are based on misinformation.  
This mechanism is extremely realistic and typical of pyramid or Ponzi scheme structures, in which, even if one is aware of participating in such a scheme, profits can still be made as long as enough new participants enter in the future.

\begin{table}[H]
    \caption{Summary of moments for different scenarios}
    \label{tab:summary}
    \centering
    \begin{tabular}{llll}
    \hline \hline
    \multicolumn{1}{c}{}     & \multicolumn{1}{c}{\textbf{Profit of Misinformed}} &  \multicolumn{1}{c}{\textbf{Skeweness}} & \multicolumn{1}{c}{\textbf{Kurtosis}} \\ \hline
    Small World              & -4.30                                                      &  0.04                 & 5.50                                  \\
    Stochastic Block Network & -3.17                                                    & 0.08                   & 7.85                                  \\
    Scale Free Informed      & 0.29                                                       & 0.04                  & 3.77                                \\
    Scale Free Misinformed   & 85.58                                                       &  -0.01                  & 5.44                                  \\ \hline \hline
    \end{tabular}
\end{table}

\section{Conclusion}\label{sec:conclusion}
We have presented an Agent-Based Model of a financial market to study the interplay between information diffusion 
and market prices and returns. In this setting agents are connected in a social network and can obtain information from 
their peers in order to form more accurate forecasts of the underlying dividend process.
We proposed a novel mechanism of expectation formation when agents have to evaluate multiple sources of news simultaneously. 
This is based on Bayesian updating and provides an alternative to the full information rational expectations assumption, while imposing minimum departures from it.
By means of numerical simulations we examined the efficiency implications of multiple social network structures. 
Leptokurtosis of returns and wealth inequality are exacerbated in societies in which misinformed agents occupy prominent positions.

Our approach in this paper is purely positive: we do not address the reasons why certain network topologies emerge, yet this work yields valuable insights for both retail investors and regulators. 
Our results indicate that in societies where misinformed agents hold prominent positions in terms of communication power, relying solely on past forecast accuracy to filter out misinformation may be insufficient. 
Although total welfare remains unchanged across different topologies, given the zero-sum nature of this simplified market, certain configurations may still be preferable, 
particularly those that mitigate wealth inequality. 
In this context, policymakers might consider promoting social media structures that prioritize friend-based connections rather than follower-based ones. 
Likewise, retail investors should try to assess the topology of the network in which they are embedded before deciding whether to incorporate a source of information in their investment strategy.

\section*{Code Availability Statement}
All data used in the paper is publicly available. 
The code used to generate the results is available at \href{https://github.com/danielTorren/Misinformation_financial_markets}{https://github.com/danielTorren/Misinformation_financial_markets}.
\medskip

\bibliography{references.bib} 

\newpage

\appendix

\begin{center}
    \Large{Appendix for \\
    \textbf{(Mis)information diffusion and the financial market}}
\end{center}

\section{Forward Looking Price}\label{sec:flp}
\begin{equation}
\begin{aligned}
p_t & =R^{-1} \Tilde{\mathbb{E}}_{t}\left(p_{t+1}+d_{t+1}\right) \\
& =R^{-1}\left[\Tilde{\mathbb{E}}_{t}\left(p_{t+1}\right)+\Tilde{\mathbb{E}}_{t}\left(d_{t+1}\right)\right] \\
& = R^{-1}\left[\Tilde{\mathbb{E}}_{t}\left(R^{-1} \Tilde{\mathbb{E}}_{t+1}\left(p_{t+2}+d_{t+2}\right)\right)+\Tilde{\mathbb{E}}_{t}\left(d_{t+1}\right)\right] \\
& = R^{-1} \Tilde{\mathbb{E}}_{t}(d_{t+1}) + R^{-2} \Tilde{\mathbb{E}}_{t}(d_{t+2}) + R^{-2} \Tilde{\mathbb{E}}_{t}(p_{t+2}) \\
& = R^{-1} \Tilde{\mathbb{E}}_{t}(d_{t+1}) + R^{-2} \Tilde{\mathbb{E}}_{t}(d_{t+2}) + R^{-2} \Tilde{\mathbb{E}}_{t}(R^{-1} \mathbb{E}_{t+2}\left(p_{t+3}+d_{t+3}\right)) \\
& \vdots \\
& = \sum_{j=1}^T R^{-j} \Tilde{\mathbb{E}}_{t}\left(d_{t+j}\right) + R^{-T} \Tilde{\mathbb{E}}_{t} (p_{t+T}),
\end{aligned}
\end{equation}

which implies, imposing $\lim_{T \to \infty} R^{-T} \Tilde{\mathbb{E}}_{t} (p_{t+T}) = 0$ that in the limit $T\to \infty$ 
\begin{equation}
p_t = \sum_{j=1}^{\infty} R^{-j} \Tilde{\mathbb{E}}_{t}\left(d_{t+j}\right).
\end{equation}

Now we focus on the term $\Tilde{\mathbb{E}}_{t}\left(d_{t+j}\right).$
We have that\footnote{Clearly if agents observe the realization of the stochastic component then $\Tilde{\mathbb{E}}_{t}\left(\theta_{t+1}\right) = \theta_{t+1}$. We opt to keep the notation general because this allows us to simultaneously treat also agents that do not observe the realization of this noisy component.}

\begin{equation}
    \Tilde{\mathbb{E}}_{t}\left(d_{t+1}\right)=\Tilde{\mathbb{E}}_{t}\left(d+\theta_{t+1}+\varepsilon_{t+1}\right)=d+\Tilde{\mathbb{E}}_{t}\left(\theta_{t+1}\right)+\Tilde{\mathbb{E}}_{t}\left(\varepsilon_{t+1}\right) =  d + \Tilde{\mathbb{E}}_{t}\left(\theta_{t+1}\right),
\end{equation}
since $\Tilde{\mathbb{E}}_{t}(\varepsilon_{t+1}) = 0.$

Similarly
\begin{equation}
    \Tilde{\mathbb{E}}_{t}\left(d_{t+2}\right)=\Tilde{\mathbb{E}}_{t}\left(d+\theta_{t+2}+\varepsilon_{t+2}\right)=d+\Tilde{\mathbb{E}}_{t}\left(\theta_{t+2}\right)+\Tilde{\mathbb{E}}_{t}\left(\varepsilon_{t+2}\right) 
=  d + \beta \Tilde{\mathbb{E}}_{t}(\theta_{t+1}), 
\end{equation}

since $\Tilde{\mathbb{E}}_{t}(\varepsilon_{t+2}) = 0$ and $\Tilde{\mathbb{E}}_{t}(\theta_{t+2}) = \Tilde{\mathbb{E}}_{t}(\beta \theta_{t+1} + \eta_{t+2}),$
and in general 
\begin{equation}
    \Tilde{\mathbb{E}}_{t}\left(d_{t+j}\right)
=  d +  \beta^{j-1} \Tilde{\mathbb{E}}_{t}(\theta_{t+1}) \quad \textrm{for all } \quad j \geq 1. 
\end{equation}

Therefore we have 

\begin{equation}
    p_t = \sum_{j=1}^{\infty} R^{-j} \left( d + \beta^{j-1} \Tilde{\mathbb{E}}_{t} (\theta_{t+1}) \right) 
= d \sum_{j=1}^{\infty} R^{-j} + \beta^{-1} \Tilde{\mathbb{E}}_{t} (\theta_{t+1}) \sum_{j=1}^{\infty} \left( \frac{\beta}{R} \right)^j.
\end{equation}
Now we have 
\begin{equation}
    d \sum_{j=1}^{\infty} R^{-j} = d \sum_{j=0}^{\infty} R^{-j} -d = d\frac{R}{R-1} -d = \frac{d}{r},
\end{equation}
since $|R^{-1}|<1$. Similarly 
\begin{equation}
    \begin{aligned}
    \beta^{-1}\Tilde{\mathbb{E}}_{t}\left(\theta_{t+1}\right) \sum_{j=1}^{\infty} \left(\frac{\beta}{R}\right)^j = \beta^{-1}\Tilde{\mathbb{E}}_{t}\left(\theta_{t+1}\right) \sum_{j=0}^{\infty} \left(\frac{\beta}{R}\right)^j - \beta^{-1}\Tilde{\mathbb{E}}_{t}\left(\theta_{t+1}\right) =
\\
= \beta^{-1}\Tilde{\mathbb{E}}_{t}\left(\theta_{t+1}\right) \frac{R}{R-\beta} - \beta^{-1}\Tilde{\mathbb{E}}_{t}\left(\theta_{t+1}\right) = \beta^{-1}\Tilde{\mathbb{E}}_{t}\left(\theta_{t+1}\right) \frac{\beta}{R-\beta}.
\end{aligned}
\end{equation}
So that finally 

\begin{equation}
    p_t = \frac{d}{r} +  \frac{\Tilde{\mathbb{E}}_{t}\left(\theta_{t+1}\right)}{R - \beta}.
\end{equation}

\section{Relationship of updating to Kalman Filter}\label{sec:relationship_kalman_filter}

Following the notation in chapter 13 of \cite{hamilton1994series} we have the following state space representation for the observable component of dividends $\theta_{t}$:

\begin{equation}
    \begin{split}
        \textrm{(state equation)} \quad & \xi_{t} = F \xi_{t-1} + v_{t},  \\
        \textrm{(measurement equation)} \quad & y_{t} = H\xi_{t} + w_{t} ,
    \end{split}
\end{equation}

in which all quantities are scalars, $\xi_{t} \equiv \theta_{t+1}$, $F\equiv\beta$, $H\equiv1$. 
The variance-covariance matrix $R$ associated with $w_{t}$ is the scalar $\sigma^2_{j,t}$. 
In this contest the a priori variance covariance matrix is given by
\begin{equation}
    P_{t|t-1} = \mathbb{E}(\xi_{t} - \hat{\xi}_{t|t-1})^2 = \sigma^2_{\eta},
\end{equation}
and the Kalman Gain 
\begin{equation}
    K_t = P_{t|t-1}H(H'P_{t|t-1}H + R)^{-1}, 
\end{equation}
collapses to 
\begin{equation}
K_t = \frac{\sigma^2_{\eta}}{\sigma^2_{\eta} + \sigma^2_{j,t}},
\end{equation}
which is exactly the weight associated to the information received by source $j$ in the case of being connected to source $j$ only.

\section{Derivation of conditional variance}\label{sec:conditional_variance}

\begin{equation}
    \Tilde{\mathbb{V}}_t(p_{t+1} + d_{t+1}) = \Tilde{\mathbb{V}}_t(d_{t+1}) + \Tilde{\mathbb{V}}_t(p_{t+1}) + 2\Tilde{\mathbb{COV}}_t(d_{t+1}, p_{t+1}).
\end{equation}
The conditional variance of next period dividends is given by

\begin{equation}
    \Tilde{\mathbb{V}}_t(d_{t+1}) =  \Tilde{\mathbb{V}}_t(\theta_{t+1}) + \sigma^2_{\varepsilon}.
\end{equation}
The conditional variance of next period price can be derived starting from the expression for $p_{t+1}$ implied by equation (\ref{price_hom}) 

We get

\begin{equation}
    \Tilde{\mathbb{V}}_t(p_{t+1}) =  \Tilde{\mathbb{V}}_t\left(\frac{d}{r} +  \frac{\Tilde{\mathbb{E}}_{t+1}(\theta_{t+2})}{R-\beta}\right) = \frac{\Tilde{\mathbb{V}}_t(\Tilde{\mathbb{E}}_{t+1}(\theta_{t+2}))}{(R-\beta)^2}.
\end{equation}

In a similar fashion we can derive the expression for the covariance of the future price and dividend as

\begin{equation}
    2\Tilde{\mathbb{COV}}_t(d_{t+1}, p_{t+1}) = 2\Tilde{\mathbb{COV}}_t\left(\theta_{t+1},  \frac{\Tilde{\mathbb{E}}_{t+1}(\theta_{t+2})}{R-\beta}\right),
\end{equation}
since $d$ is a constant and $\varepsilon_{t+1}$ is independent of any other variable.
Rearranging we get equation (\ref{eq:exp_variance}).

\section{Proofs}

\subsection{Proof of Proposition(\ref{thm:BA})}\label{sec:proof_BA}

The proof is by induction. First we prove the statement for the base case with only one source, that is $\Bar A = 2$.
This collapses to the normal case of Bayesian updating with conjugate normal prior, therefore we have:

\begin{equation}
\mu_P = \frac{\mu_1\sigma^2_0 + \mu_0\sigma^2_1}{\sigma^2_0 + \sigma^2_1},
\end{equation}

\begin{equation}
\sigma^2_P = \frac{\sigma^2_0\sigma^2_1}{\sigma^2_0 + \sigma^2_1}.
\end{equation}

Then we prove that if the statement holds for a generic set A with $\Bar A = n$, then it holds also for B with $\Bar B = n+1$.
If the statement holds for $\Bar A = n$, and receive an extra source of information, the new posterior distribution 
has parameters:

\begin{equation}
\mu_{P} = \frac{\mu_{K+1}\frac{\prod^{K}_{j=0}\sigma^2_j}{\sum \left[ A \right]^{\Bar{A}-1}} + \frac{\sum^{K}_{k=0}\left(\mu_k \cdot \left[ A \right]^{\Bar{A}-1}_{k}\right)}{\sum \left[ A \right]^{\Bar{A}-1}}\sigma^2_{K+1}}{\frac{\prod^{K}_{j=0}\sigma^2_j}{\sum \left[ A \right]^{\Bar{A}-1}} + \sigma^2_{K+1}} 
= \frac{ \frac{\sum^{K+1}_{k=0}\left(\mu_k \cdot \left[ A \cup \sigma^2_{K+1}\right]^{\Bar A}_{k}\right)}{\sum \left[ A \right]^{\Bar{A}-1}}}{\frac{\prod^{K}_{j=0}\sigma^2_j}{\sum \left[ A \right]^{\Bar{A}-1}} + \frac{\sum \left[\left[ A \cup  \sigma^2_{K+1}\right]^{\Bar{A}} \setminus\prod^{K}_{j=0}\sigma^2_j \right]}{\sum \left[ A \right]^{\Bar{A}-1}}} = 
\end{equation}

\begin{equation}
 =  \frac{\sum^{K+1}_{k=0}\left(\mu_k \cdot \left[ A \cup \sigma^2_{K+1}\right]^{\Bar A}_{k}\right)}{\sum\left[ A \cup  \sigma^2_{K+1}\right]^{\Bar{A}}}.
\end{equation}

\begin{equation*}
\sigma^2_{P} = \frac{\frac{\prod^{K}_{j=0}\sigma^2_j}{\sum \left[ A \right]^{\Bar{A}-1}}\sigma^2_{K+1}}{\frac{\prod^{K}_{j=0}\sigma^2_j}{\sum \left[ A \right]^{\Bar{A}-1}} + \sigma^2_{K+1}} =  \frac{\frac{\prod^{K+1}_{j=0}\sigma^2_j}{\sum \left[ A \right]^{\Bar{A}-1}}}{\frac{\prod^{K}_{j=0}\sigma^2_j}{\sum \left[ A \right]^{\Bar{A}-1}} + \frac{\sum \left[\left[ A \cup  \sigma^2_{K+1}\right]^{\Bar{A}} \setminus\prod^{K}_{j=0}\sigma^2_j \right]}{\sum \left[ A \right]^{\Bar{A}-1}}} = \frac{\prod^{K+1}_{j=0}\sigma^2_j}{\sum\left[ A \cup  \sigma^2_{K+1}\right]^{\Bar{A}}}.
\end{equation*}

Therefore we have:

\begin{equation}
\mu_{P} = \frac{\sum^{K+1}_{k=0}\left(\mu_k \cdot \left[B\right]^{\Bar B -1}_{k}\right)}{\sum\left[ B\right]^{\Bar{B}-1}},
\end{equation}

\begin{equation}
\sigma^2_{P} = \frac{\prod^{K+1}_{j=0}\sigma^2_j}{\sum\left[ B\right]^{\Bar{B}-1}},
\end{equation}

which are the posterior mean and variance for the set $B = A \cup  \sigma^2_{K+1}$ with $\Bar{B} = \Bar{A} +1 = n+1$, 
hence concluding the proof.

\subsection{Proof of Proposition (\ref{prop:van_posterior})}\label{sec:proof_van}

Start from equation (\ref{post_variance}). Defining as $e^k$ the posterior variance obtained when agent has K connections,
we can express the posterior variance recursively for increasing values of K as
\begin{equation}
    e^K = e^{K-1} \frac{\sum\left[ A \right]^{K-1} \sigma^2_K}{\sum\left[ A \right]^{K}}.
\end{equation}
To prove this formulation we can again rely on induction. The base case is $K=1$ for which 
\begin{equation}
    e^1 = \frac{\sigma^2_0\sigma^2_1}{\sigma^2_0 + \sigma^2_1}.
\end{equation}

Then if the formula holds for $K$ we have
\begin{equation}
    e^{K+1} = e^{K} \frac{\sum\left[ A \right]^{K} \sigma^2_{K+1}}{\sum\left[ A \right]^{K+1}},
\end{equation}

and iterating 
\begin{equation}
    e^{K+1} = e^{1} \frac{\sum\left[ A \right]^{1} \sigma^2_{2}}{\sum\left[ A \right]^{2}} \frac{\sum\left[ A \right]^{2} \sigma^2_{3}}{\sum\left[ A \right]^{3}} \dots \frac{\sum\left[ A \right]^{K} \sigma^2_{K+1}}{\sum\left[ A \right]^{K+1}},
\end{equation}
ans substituting the formula for $e^1$ by noticing that $\sum\left[ A \right]^{1} = \sigma^2_0 + \sigma^2_1$ we have 
\begin{equation}
    e^{K+1} =  \frac{\sigma^2_0\sigma^2_1}{\sum\left[ A \right]^{1}} \frac{\sum\left[ A \right]^{1} \sigma^2_{2}}{\sum\left[ A \right]^{2}} \frac{\sum\left[ A \right]^{2} \sigma^2_{3}}{\sum\left[ A \right]^{3}} \dots \frac{\sum\left[ A \right]^{K} \sigma^2_{K+1}}{\sum\left[ A \right]^{K+1}},
\end{equation}

which simplifying gives exactly 

\begin{equation}
    e^{K+1} = \frac{\prod^{K+1}_{j=0}\sigma^2_j}{\sum \left[ A \right]^{K+1}}.
\end{equation}

The reason why the recursive formula is useful is that 
\begin{equation}
    \frac{\sum\left[ A \right]^{K} \sigma^2_{K+1}}{\sum\left[ A \right]^{K+1}} < 1,
\end{equation}

that is

\begin{equation}
    \sum\left[ A \right]^{K} \sigma^2_{K+1} <{\sum\left[ A \right]^{K+1}} ,
\end{equation}

which holds noticing that $ {\sum\left[ A \right]^{K+1}} - \sum\left[ A \right]^{K} \sigma^2_{K+1} = \prod^K_{j=0} \sigma^2_j$ and $\sigma^2_j > 0 \forall j.$
This combined with the fact that the variances are bounded makes so that \begin{equation}
    \lim_{K \to \infty} e^K = 0.
\end{equation}

\section{Validation of the SNPE}\label{sec:SNPE_validation}  
\begin{figure}[H]
    \centering
    \includegraphics[width=0.6\textwidth]{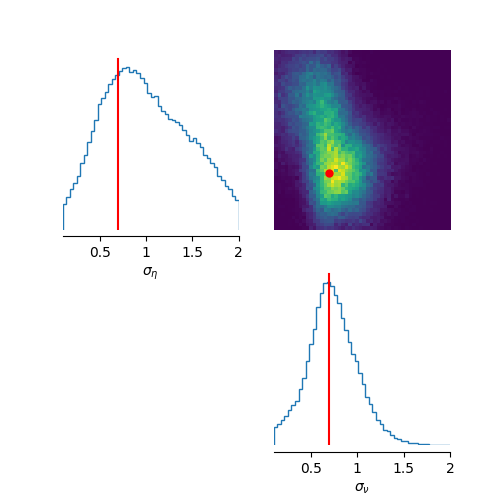}
    \caption{Validation of the SNPE. }
    \label{fig:SNPE_validation}
    \small{Note: The Figure shows the posterior distribution of the parameters obtained by the SNPE algorithm. The red lines represent the true values of the parameters.}
\end{figure}

To validate the model we generate synthetic data from the ABM and then we use the SNPE algorithm to recover the parameters.
We fix $\sigma_{\eta} = \sigma_{\nu} = 0.7$ and keep the other parameters to their values in table \ref{tab:param_calib}.
We choose a uniform prior for the parameters with support $[0.1, 2]$ and we provide the result of 10000 stochastic simulations to train the density estimator.
The result is displayed in Figure \ref{fig:SNPE_validation} and we can see that the posterior distribution is extremely weel behaved
with median values $0.95$ and $0.71$ respectively for $\sigma_{\eta}$ and $\sigma_{\nu}$.
The value of the first parametere is slightly off, but well within the confidence interval, and moslty likely due the fact that the 
posterior looks flat in the region of the true value. 

\end{document}